\documentclass[10pt,twocolumn,letterpaper]{article}
\usepackage{usenix}
\usepackage{url}
\usepackage{adjustbox}
\usepackage{amssymb}
\usepackage{amsmath}
\usepackage{amsthm}
\usepackage{anyfontsize}
\usepackage{booktabs}
\usepackage[font={small}]{caption}
\usepackage{color}
\usepackage{colortbl}
\usepackage{enumitem}
\usepackage{framed}
\PassOptionsToPackage{hyphens}{url}
\usepackage[hidelinks,bookmarks=false]{hyperref}
\usepackage[numbers,sort]{natbib}
\usepackage[letterpaper,left=0.65in,right=0.65in,top=0.75in,bottom=0.75in]{geometry}
\usepackage{mathtools}
\usepackage{microtype}
\usepackage{scalefnt}
\usepackage{setspace}
\usepackage{tikz}
\usetikzlibrary{calc}
\usepackage{xcolor}
\usepackage{xspace}
\usepackage{titlesec}
\usepackage{wrapfig}

\newcommand{\id}{\mathsf{id}}
\newcommand{\overflow}{\mathsf{overflow}}
\newcommand{\F}{\mathbb{F}}

\newcommand{\Z}{\mathbb{Z}}

\newcommand{\G}{\mathbb{G}}

\newcommand{\VRFKeyGen}{\mathsf{VRF.KeyGen}}
\newcommand{\VRFEval}{\mathsf{VRF.Eval}}
\newcommand{\VRFVerify}{\mathsf{VRF.Verify}}

\newcommand{\VIFKeyGen}{\mathsf{VIF.KeyGen}}
\newcommand{\VIFEval}{\mathsf{VIF.Eval}}
\newcommand{\VIFVerify}{\mathsf{VIF.Verify}}

\newcommand{\SigKeyGen}{\mathsf{Sig.KeyGen}}
\newcommand{\SigSign}{\mathsf{Sig.Sign}}
\newcommand{\SigVerify}{\mathsf{Sig.Verify}}

\newcommand{\EKeyGen}{\mathsf{ECDSA.KeyGen}}
\newcommand{\ESign}{\mathsf{ECDSA.Sign}}
\newcommand{\EVerify}{\mathsf{ECDSA.Verify}}

\newcommand{\Init}{\mathsf{Count.Init}}

\newcommand{\Inc}{\mathsf{Count.Inc}}

\newcommand{\calA}{\ensuremath{\mathcal{A}}}
\newcommand{\calB}{\ensuremath{\mathcal{B}}}

\newcommand{\calD}{\ensuremath{\mathcal{D}}}

\newcommand{\calF}{\ensuremath{\mathcal{F}}}
\newcommand{\calG}{\ensuremath{\mathcal{G}}}

\newcommand{\calI}{\ensuremath{\mathcal{I}}}

\newcommand{\calM}{\ensuremath{\mathcal{M}}}

\newcommand{\calO}{\ensuremath{\mathcal{O}}}

\newcommand{\calS}{\ensuremath{\mathcal{S}}}

\newcommand{\calV}{\ensuremath{\mathcal{V}}}

\newcommand{\calX}{\ensuremath{\mathcal{X}}}
\newcommand{\calY}{\ensuremath{\mathcal{Y}}}

\newcommand{\deq}{\mathrel{\mathop:}=}
\newcommand{\zo}{\ensuremath{\{0,1\}}} 

\theoremstyle{plain}
\newtheorem{theorem}{Theorem}

\newtheorem{lemma}[theorem]{Lemma}

\theoremstyle{definition}

\newtheorem{defn}[theorem]{Definition}

\newtheoremstyle{goal}
  {\topsep}
  {\topsep}
  {\normalfont}
  {0pt}
  {\bfseries}
  {: } 
  { }
  {\thmname{#1}\thmnumber{ #2}\thmnote{ (#3)}}
\theoremstyle{goal}

\newcommand{\esm}[1]{\ensuremath{#1}}
\newcommand{\ms}[1]{\esm{\mathsf{#1}}}

\newcommand{\getsr}{\rgets}
\newcommand{\rgets}{\mathrel{\mathpalette\rgetscmd\relax}}
\newcommand{\rgetscmd}{\ooalign{$\leftarrow$\cr
    \hidewidth\raisebox{1.2\height}{\scalebox{0.5}{\ \rm R}}\hidewidth\cr}}

\newcommand{\abs}[1]{\left| #1 \right|}

\newcommand{\View}{\ms{View}}
\newcommand{\Sim}{\ms{Sim}}

\newcommand{\st}{\ms{st}}

\newcommand{\sk}{\ms{sk}}
\newcommand{\pk}{\ms{pk}}

\newcommand{\skVRF}{\ms{sk}_\mathsf{VRF}}
\newcommand{\pkVRF}{\ms{pk}_\mathsf{VRF}}
\newcommand{\piVRF}{\pi_\mathsf{VRF}}

\newcommand{\msk}{\ms{msk}}
\newcommand{\mpk}{\ms{mpk}}

\newcommand{\UFAdv}{\ms{VIF\textsf{-}UFAdv}}
\newcommand{\UFAdvs}{\ms{VIF\textsf{-}UFAdv}^\ms{single}}
\newcommand{\DLAdv}{\ms{DlogAdv}}
\newcommand{\VRFAdv}{\ms{VRF\textsf{-}PRAdv}}
\newcommand{\VRFSetAdv}{\ms{VRF}\textsf{-}\ms{SetAdv}}
\newcommand{\PRAdv}{\ms{VIF\textsf{-}PRAdv}}
\newcommand{\Fdsa}{\calF^{\G,\calV}_{\text{ECDSA}}}
\newcommand{\Bvrf}{\calB_\mathsf{VRF}}
\newcommand{\Bdlog}{\calB_\mathsf{Dlog}}

\newcommand{\Qid}{Q_\ms{id}}

\newcommand{\Qsig}{Q_\ms{sig}}

\newcounter{ExperimentCount}
\newcommand{\Experiment}[1]{\refstepcounter{ExperimentCount}\textbf{Experiment~\arabic{ExperimentCount}: #1.}}

\newcommand{\Wz}[1]{p_{\ref{#1},0}}
\newcommand{\Wo}[1]{p_{\ref{#1},1}}
\newcommand{\Wb}[1]{p_{\ref{#1},b}}
\newcommand{\Win}[1]{p_{\ref{#1}}}

\newcommand{\ID}{\mathcal{I}}
\newcommand{\Dreal}{\calD_\text{real}}
\newcommand{\Dideal}{\calD_\text{ideal}}

\newcommand{\Oid}{\calO_\ms{id}}
\newcommand{\Osig}{\calO_\ms{sig}}

\newcommand{\wR}{R_\textrm{abs}}

\newcommand{\name}{True2F\xspace}

\newcommand{\figureArrow}[4]{
\multicolumn{3}{c}{
  \begin{picture}(80,20)
  \put(40,8){\makebox(0,0){#1}}
  \put(#2,0){\vector(#3,#4){80}}
  \end{picture}
}\vspace{-10pt}}

\newcommand{\figureArrowLR}[1]{
\multicolumn{3}{c}{
  \begin{picture}(80,20)
  \put(40,8){\makebox(0,0){#1}}
  \put(0,0){\vector(1,0){80}}
  \put(80,0){\vector(-1,0){80}}
  \end{picture}
}\vspace{-10pt}}

\newcommand{\figureRightArrow}[1]{\figureArrow{#1}{0}{1}{0}}
\newcommand{\figureLeftArrow}[1]{\figureArrow{#1}{80}{-1}{0}}
\newcommand{\figureBothArrow}[1]{\figureArrowLR{#1}}
\newcommand{\appID}{\ms{appID}}
\newcommand{\chal}{\ms{chal}}

\newcommand{\batch}{\ms{att}}
\newcommand{\keys}{\ms{Keys}}

\newcommand{\stD}{\ms{st}}
\newcommand{\stB}{\ms{st}'}

\newcommand{\eabort}{\epsilon_\ms{abort}}
\newcommand{\ttO}{\texttt{1}\xspace}
\newcommand{\ttZ}{\texttt{0}\xspace}

\newcommand{\Atoken}{\calA_\ms{t}}
\newcommand{\Arp}{\calA_\ms{rp}}

\newcommand{\NameAuthMs}{57\xspace}
\newcommand{\NameRegMs}{109\xspace}

\newcommand{\UAuthMs}{23\xspace}
\newcommand{\URegMs}{64\xspace}

\makeatletter
\g@addto@macro{\UrlBreaks}{\UrlOrds}
\makeatother

\usepackage[T1]{fontenc}
\usepackage{newtxmath}
\usepackage{newtxtext}

\usepackage[scaled=.8]{beramono}

\DeclareMathAlphabet\mathcal{OMS}{cmsy}{m}{n}
\SetMathAlphabet\mathcal{bold}{OMS}{cmsy}{b}{n}

\DeclareSymbolFont{AMSb}{U}{msb}{m}{n}
\DeclareSymbolFontAlphabet{\mathbb}{AMSb}

\DeclareSymbolFont{numbers}{T1}{ptm}{m}{n}
\SetSymbolFont{numbers}{bold}{T1}{ptm}{bx}{n}
\DeclareMathSymbol{0}\mathalpha{numbers}{"30}
\DeclareMathSymbol{1}\mathalpha{numbers}{"31}
\DeclareMathSymbol{2}\mathalpha{numbers}{"32}
\DeclareMathSymbol{3}\mathalpha{numbers}{"33}
\DeclareMathSymbol{4}\mathalpha{numbers}{"34}
\DeclareMathSymbol{5}\mathalpha{numbers}{"35}
\DeclareMathSymbol{6}\mathalpha{numbers}{"36}
\DeclareMathSymbol{7}\mathalpha{numbers}{"37}
\DeclareMathSymbol{8}\mathalpha{numbers}{"38}
\DeclareMathSymbol{9}\mathalpha{numbers}{"39}

\SetSymbolFont{numbers}{bold}{T1}{ptm}{b}{n}
\makeatletter
\renewcommand{\operator@font}{\mathgroup\symnumbers}
\makeatother

\makeatletter

\renewcommand\small{%
  \@setfontsize\small{9pt}{10pt}%
   \abovedisplayskip 8.5\p@ \@plus3\p@ \@minus4\p@
   \abovedisplayshortskip \z@ \@plus2\p@
   \belowdisplayshortskip 4\p@ \@plus2\p@ \@minus2\p@
   \def\@listi{\leftmargin\leftmargini
               \topsep 4\p@ \@plus2\p@ \@minus2\p@
               \parsep 2\p@ \@plus\p@ \@minus\p@
               \itemsep \parsep}%
   \belowdisplayskip \abovedisplayskip
}
\makeatother

\definecolor{LightCyan}{rgb}{0.88,1,1}

\long\def\com#1{}

\newcommand{\itpara}[1]{\medskip\noindent\textit{#1}}
\newcommand{\sitpara}[1]{\smallskip\noindent\textit{#1}}
\renewcommand{\paragraph}[1]{\medskip\noindent\textbf{#1}}

\makeatletter

\let\c@table\c@figure
\makeatother 

\setlist[description]{leftmargin=\parindent,topsep=0ex,itemsep=0ex,partopsep=1ex,parsep=1ex}

\LetLtxMacro{\oldtextsc}{\textsc}
\renewcommand{\textsc}[1]{\oldtextsc{\scalefont{1.2}#1}}

\setitemize{noitemsep,topsep=2pt,parsep=2pt,partopsep=2pt,leftmargin=4ex}
\setenumerate{noitemsep,topsep=2pt,parsep=2pt,partopsep=2pt,leftmargin=4ex}

\newcolumntype{R}[2]{%
  >{\adjustbox{angle=#1,lap=\width-(#2)}\bgroup}%
    c%
    <{\egroup}%
}

\newcommand{\rot}[1]{{\adjustbox{angle=60,lap=\width-1em}{#1}}}
\newcommand{\yrot}[1]{{%
  \begingroup \setlength\fboxsep{2pt}
  \adjustbox{angle=60,lap=\width-1em}{\colorbox{yellow!30}{#1}}%
  \endgroup}}

\begin{document}

\title{True2F: Backdoor-resistant authentication tokens\\
{\normalsize {\normalfont \today}}} 
\author{Emma Dauterman\\
\textit{Stanford and Google}
\and
Henry Corrigan-Gibbs\\
\textit{Stanford}
\and
David Mazi\`eres\\
\textit{Stanford}
\and
Dan Boneh\\
\textit{Stanford}
\and
Dominic Rizzo\\
\textit{Google}
}
\maketitle

\frenchspacing
\paragraph{Abstract.}
We present \name, 
a system for second-factor authentication that provides the 
benefits of conventional authentication tokens
in the face of phishing and software compromise, 
while also providing strong protection against token faults and backdoors. 
To do so, we develop new lightweight two-party protocols for 
generating cryptographic keys and ECDSA signatures, and we implement new
privacy defenses to prevent cross-origin token-fingerprinting attacks.
To facilitate real-world deployment,
our system is backwards-compatible with today's U2F-enabled web services 
and runs on commodity hardware tokens after a firmware modification.
A \name-protected authentication takes just \NameAuthMs{}ms
to complete on the token, compared with \UAuthMs{}ms
for unprotected U2F.

\section{Introduction}
Two-factor authentication has become a standard defense against
weak passwords, keyloggers, and other types of malware.
Universal Second Factor (U2F) hardware authentication tokens are an especially
effective
type of second-factor authenticator.
Because these tokens cryptographically bind their authentication
messages to a specific origin,
they block phishing attacks to
which other approaches, such as time-based one-time
passwords (e.g., Google Authenticator), leave users exposed.  
These tokens run on simple, dedicated hardware, and so they present a much
smaller attack surface than mobile-phone-based methods do. 
Since Google mandated in early 2017 that all of its employees use 
U2F authentication tokens, the company has not
discovered a single instance of corporate credential theft~\cite{krebs}.

But the introduction of new, and often opaque,
hardware components into a system carries additional risks as well.
If these components are poorly designed or even intentionally backdoored, 
they can \textit{undermine} the security of an 
otherwise well-functioning system. 
A now-infamous bug in the design of Infineon-made cryptographic chips led to
critical vulnerabilities in millions of TPMs, electronic passports, laptops,
and hardware tokens~\cite{ROCA17}.
Users of these flawed Infineon chips ended up \textit{worse off} than those who just
implemented their cryptographic routines using standard software libraries.

The consequences of faulty or backdoored hardware authentication
tokens can be equally dire. 
If a token's randomness source is faulty---a common failure mode for
low-cost devices~\cite{PQ12,RonWhit,Ever14,Y09}---then 
anyone who phishes a user's credentials can authenticate as that
user, even without access to the user's machine.
An attacker who compromises a hardware token during manufacturing
or via supply-chain tampering~\cite{supplychain,feitian,feitian2,wallet-reseller} can exfiltrate
the user's cryptographic credentials or authentication history 
(e.g., to an attacker-controlled web service).
The latest batch of web standards~\cite{fido2,webauthn,webauthn-press} allow using
hardware tokens for password-less ``first-factor'' authentication, which only
exacerbates these dangers.

This raises the question:
Can we enjoy the \textit{benefits} of hardware-level
security protections without incurring the \textit{risks} of introducing
additional (and often untrustworthy) hardware components into our systems?
In this paper, we give a strong positive answer to this question
for the interesting special case of hardware authentication tokens.

\medskip

We present the design and implementation of \name, a system
for two-factor authentication that simultaneously provides
the benefits of conventional second-factor authentication hardware tokens and
very strong protection against the risks of token faults or backdoors.
\name requires only modest modifications to the token firmware and browser software.
Critical for enabling incremental deployment is that \name runs on 
\textit{unmodified token hardware} and is backwards compatible with
online services that already support U2F; using \name 
requires \textit{no server-side changes whatsoever}.

\name uses the same authentication process as U2F: a relying party (i.e., web
service) sends a challenge to the user's browser, the browser forwards the
challenge to a hardware token via USB, and the token returns a signature on the
challenge to the relying party via the browser.
The only difference is that a \name-compliant token and browser exchange
a few extra messages before responding to the relying party.
These messages allow the browser to enforce the token's correct
behavior, preventing a malicious token from choosing weak or
preloaded keys, or from covertly leaking keying material in messages
to the relying party.

\begin{figure}
  \centering
\begin{minipage}[c]{0.475\columnwidth}
\includegraphics[width=\columnwidth]{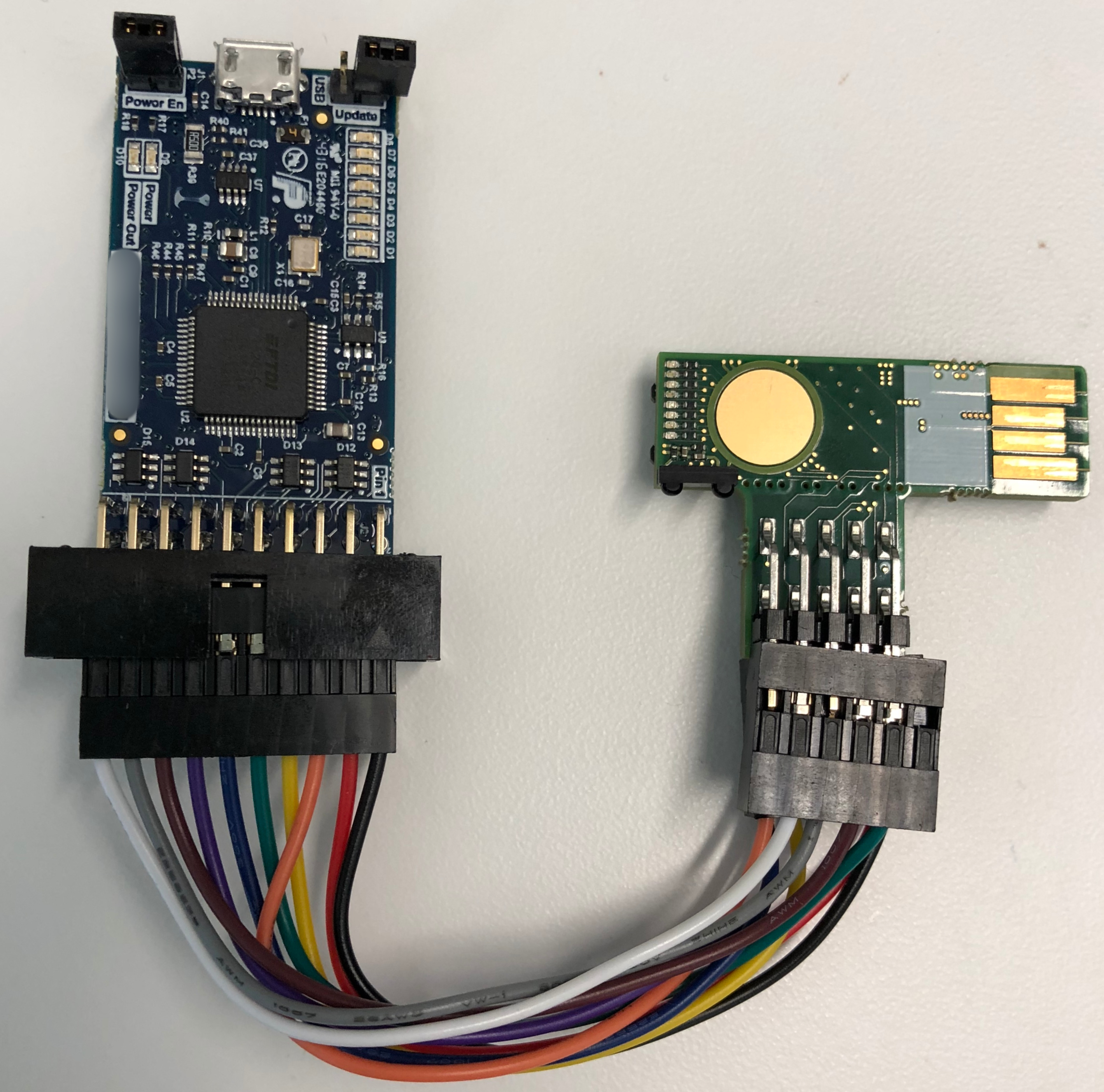}
\end{minipage}\hspace{0.02\columnwidth}%
\begin{minipage}[t]{0.475\columnwidth}
\includegraphics[width=\columnwidth]{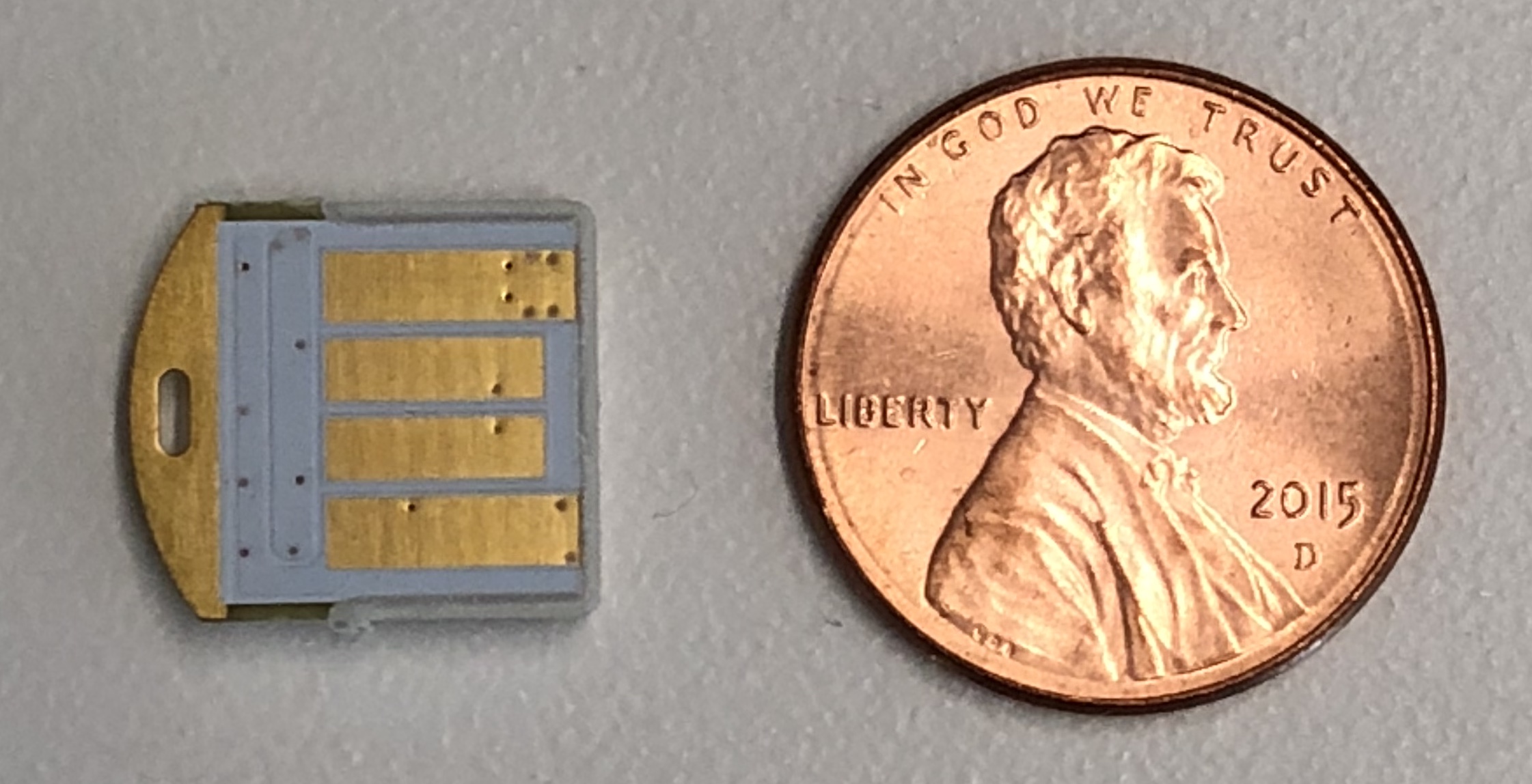}
  \caption{A development board that runs our \name prototype (at left) and
  a production USB token capable of running \name (above).}\label{fig:hw}
\end{minipage}
\end{figure}

The \name token provides the same level of protection against phishing
and browser compromise as a standard U2F hardware authentication token does.
After the user has registered the token with a relying party, even if the
attacker takes control of the user's machine, the attacker still cannot
authenticate to the relying party without interacting with the token.
This holds even if the attacker can passively observe the browser's 
interactions with the token before taking control of the machine.

\name additionally protects against token backdoors.
If the browser executes the \name protocol correctly, then even if the 
adversary controls a coalition of relying parties, the relying parties
cannot detect whether they are interacting with an adversarial U2F token 
or an ideal (honest) U2F token.
In particular, the token cannot exfiltrate data to the outside world via 
U2F protocol messages.

With \name, to compromise the cryptographic keys on the token,
an attacker has to compromise \textit{both} the user's machine
and the hardware token itself.
In this way, \name provides ``strongest-link'' security, while
standard U2F provides security that is only as good as the 
hardware token itself.

\medskip

The design of \name takes advantage of the fact that U2F hardware tokens
perform only very simple computations:
a U2F token generates keypairs for the ECDSA digital signature scheme,
it signs server-provided challenges, and it keeps a counter to 
track how many authentications it has performed.

We develop lightweight two-party protocols that allow the token
and browser to collaboratively execute all of these operations in
such a way that (1) the browser learns nothing about the token's secrets
and yet (2) the browser can enforce the token's strict compliance
with the U2F specification.
Our security definitions draw on the recent theoretical work on cryptographic
firewalls~\cite{MSD15,DMS16} and algorithm-substitution attacks~\cite{BPR14}.

In particular, we develop a new \textit{collaborative key-generation protocol}
that runs between the browser and token.
The browser uses this protocol to ensure that the token's master keypair
incorporates proper randomness.
Our protocol is inspired by, but is roughly $3\times$ faster
than, a scheme of prior work~\cite{randkeys}.

We introduce \textit{verifiable identity families}, a
mechanism for deterministically deriving 
an exponential number of random-looking ECDSA
keypairs from a single master keypair.
We use this primitive in \name to derive the per-website 
keypairs used for U2F authentication in a deterministic and
browser-auditable way.

We also construct \textit{firewalled ECDSA signatures},
which allow the browser and token to jointly produce ECDSA signatures on 
messages 
(1) without revealing any of the token's secrets to the browser and 
(2) while preventing the token from exfiltrating secrets via the bits of
the signature.
This prevents an adversarial token from tampering with the randomness
used to generate the signatures and from encoding bits of secret
information (e.g., the token's signing keys) in the randomness
of the signature itself~\cite{YY97}.
The innovation is that our protocol outputs unmodified ECDSA
signatures, which is critical for backwards compatibility.
In Section~\ref{sec:rel}, we explain how our construction
relates to \textit{subliminal-free}~\cite{Simm84,Simm85,Des94,Des90,D88} 
and \textit{subversion-resistant} signature schemes~\cite{AMV15}.

Finally, we implement a \textit{flash-friendly counting data structure}
that we use to implement the U2F authentication counters.
As we will describe, using finer-grained counters better protects
the user from cross-site token-fingerprinting attacks.
To make it practical to keep fine-grained counters, we propose a
new log-structured counter design that respects the tight design and space
constraints of our token's flash hardware.

\medskip

We have implemented the full \name protocol and run it
on a hardware token used by Google employees (Figure~\ref{fig:hw}).
On this token, a \name authentication completes
in \NameAuthMs{}ms, compared with \UAuthMs{}ms for a standard U2F authentication.
Registering the \name token with a new website takes \NameRegMs{}ms,
compared with \URegMs{}ms for a standard U2F registration.

\paragraph{Secondary application: Hardware wallets.} 
While our focus is on authentication tokens, 
the principles of \name also apply to protecting
cryptocurrency ``hardware wallets'' against
backdoors~\cite{wallets,wallet-reseller}.  
Hardware wallets perform essentially the same operations
as U2F tokens: ECDSA key generation and signing.
When using \name in the context of hardware wallets, the user would 
store the token's master secret key in cold storage for backup.
Otherwise, the application to hardware wallets is almost immediate, 
and we do not discuss it further.

\section{Background}
\label{sec:bg}

The primary goal of the Universal Second Factor (U2F)~\cite{u2f,LCBS} standard is to
defeat credential-phishing attacks~\cite{hong2012state} by requiring users of web
services to authenticate not only with a human-memorable password (``the first
factor'') but also using a cryptographic secret stored on a hardware token (``a
second factor'').
Even if the user accidentally discloses their login credentials to an attacker, the
attacker cannot access the user's account without also having physical access 
to the hardware token.
All major web browsers support U2F, as do many
popular web services such as Gmail, Facebook, Dropbox, and
Github~\cite{yubico-u2f,webauthn,webauthn-press,fido2}.

Three entities make up a U2F deployment:
\begin{enumerate}
  \item \textit{a hardware token}, which is a small USB device, as in Figure~\ref{fig:hw}, 
          that stores cryptographic keys and computes digital signatures, 
  \item \textit{a web browser} that serves as the user's interface to the token
      and the web, and 
  \item \textit{a relying party} (i.e., service) to which the user wants to
  authenticate.
\end{enumerate}

The U2F protocol specifies two actions: \textit{registration}
and \textit{authentication}. 
Both actions begin when the relying party makes a U2F request to the browser
via a JavaScript API.
The browser annotates the request with connection metadata, forwards it
to the U2F token via USB, and displays a prompt to the user,
who physically touches the token to complete the request.

\begin{figure}
\includegraphics[width=\columnwidth]{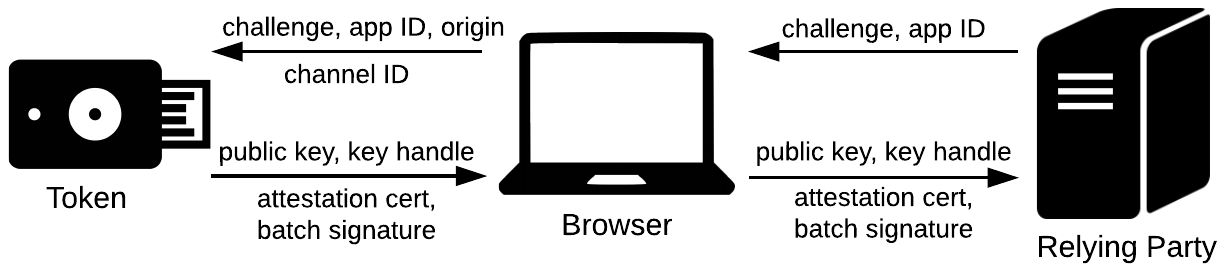}
\includegraphics[width=\columnwidth]{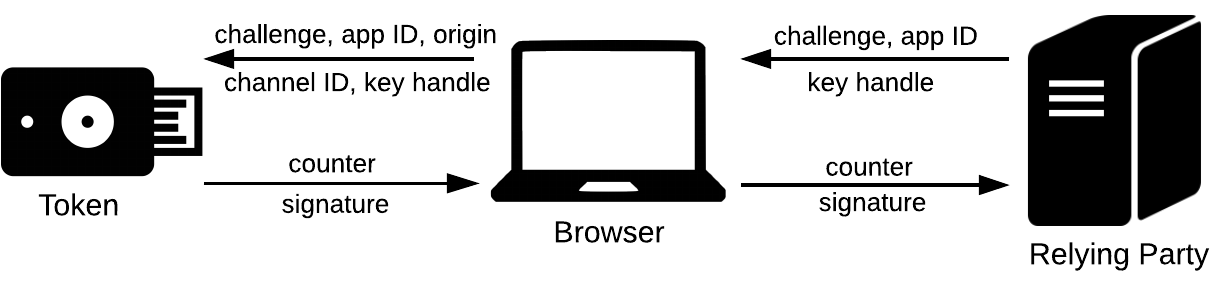}
\caption{U2F registration (top) and authentication (bottom).}
\label{fig:u2f}
\end{figure}

\paragraph{Registration.} 
Registration associates a particular hardware token with a user's account at
a relying party (e.g., \texttt{user@github.com}).
During registration the relying party sends an application identifier 
(e.g., \texttt{github.com}) and a challenge to the token, as in Figure~\ref{fig:u2f}.
The token then generates a ``per-identity'' ECDSA keypair $(\sk_\id, \pk_\id)$, 
and returns the public key $\pk_\id$ to the relying party via the browser.
Since a single user may have multiple accounts with the same relying party, the
identity $\id$ corresponds to an account-site pair, such as \texttt{user@github.com}.

The token also returns an opaque blob, called the ``key handle,'' to the relying party.
The relying party stores the public key $\pk_\id$ and key handle alongside
the user's credentials.
During authentication, the relying party passes
this key handle back to the token as part of its authentication request.
The token can use the key-handle blob to reduce the amount of keying
material it needs to store.
For example, Yubico's U2F key~\cite{yubikey} derives a per-identity
secret key by applying a pseudorandom function~\cite{GGM84}, keyed with a global
secret key, to the identity's key handle.

U2F also supports a simple form of hardware attestation.
For privacy reasons, the Chrome browser essentially disables
this feature and \name follows Chrome's behavior~\cite{chrome-batch}.

\paragraph{Authentication.} 
During authentication, the relying party sends to the browser the key handle produced at
registration, the application identifier $\appID$, and a challenge for the
token to sign (Figure~\ref{fig:u2f}). 
The browser constructs a challenge with the relying party's challenge and
connection metadata, and forwards the key handle and $\appID$ to the token.
The token uses the key handle to recover the per-identity secret signing key $\sk_\id$. 
In addition, the token reads and increments a counter it stores locally.
Finally, the token uses $\sk_\id$ to sign the browser's
challenge along with the application ID and counter value.

The counter defends against physical cloning attacks.
If an attacker clones a hardware token, then the original token and the clone
will have the same counter value.
If both tokens authenticate, then the relying party can notice that the
counter has not increased (or possibly has decreased) and lock the account.
U2F does not specify how the counter must be implemented---the only requirement
is that the counter reported at each authentication attempt to a
given relying party must be strictly increasing.
Yubico's U2F key uses a single global counter~\cite{yubikey}.

\section{Design overview}
\label{sec:goals}

In this section,
we sketch the high-level design ideas behind \name
and then describe our security goals for the system.

\subsection{From a ``Straw-man'' to \name}
A simple way to prevent a backdoored hardware token from misbehaving 
would be to make its behavior deterministic: 
after purchasing the token, the user would load a single seed into the
token. 
The token would then deterministically derive all of its randomness and
cryptographic secrets from this seed using a pseudorandom generator.
If the user stored the same cryptographic seed on the browser, the 
browser could mimic the behavior of the token, and could thus detect
whether the token ever deviated from the U2F specification.

This ``straw-man'' solution has a number of critical drawbacks.
First, to be able to audit the token, the browser must store the seed. 
If an attacker ever compromises the browser, 
the attacker can steal the seed and use it to 
learn all of the token's cryptographic secrets.
Storing the seed on the browser would then
completely defeat the purpose
of having a hardware token in the first place.

We address this first issue using a new primitive that we call a
\textit{verifiable identity family} (Section~\ref{sec:crypto:vif}).
When the user initializes the token, she generates a master
keypair $(\msk, \mpk)$, consisting of a secret key and a public key.
The user loads the master secret key onto the token and
stores the master public key on the browser.
When the browser initiates the U2F registration protocol
for an identity $\id$, the token then uses the
master secret key $\msk$ to \textit{deterministically} 
generate a unique (but random-looking)
authentication keypair $(\sk_\id, \pk_\id)$. 
Using the master public key $\mpk$, the browser can verify that $\pk_\id$
really is the unique public key that corresponds to identity $\id$.
In this way, the browser can verify that the token is generating its
cryptographic keys according to our specification.

But even this solution is imperfect: if the user generates the master
keypair $(\msk, \mpk)$ on her computer, traces of the master secret key
might be left in memory or swap files on her machine~\cite{czeskis2008defeating,garfinkel2004data,chow2005shredding}.
Malware that later compromises the browser could recover 
the master secret key $\msk$ and
thereby learn all of the token's cryptographic secrets.

To address this second issue, we use a new 
\textit{collaborative key-generation protocol} to
generate the master keypair (Section~\ref{sec:crypto:keygen}).
At the end of the protocol, the token holds the master secret key and the browser
holds the master public key.
The protocol further enforces that
(a) the keypair is sampled from the correct distribution and
(b) the browser learns nothing except the master public key. 
By using this protocol for key generation, the browser can
ensure that the token uses a master key that incorporates
sufficient randomness and the token ensures the browser never sees
its cryptographic secrets.

Even these two improvements are still not quite sufficient 
to achieve a fully backdoor-resistant U2F token.
The last remaining issue is the ECDSA signature that the
token produces in response to each authentication request.
Since ECDSA signatures are randomized, a faulty token could
use bad randomness to generate the signatures.
In addition, a malicious token could leak bits of 
information (e.g., about its master secret key) by hiding them in the
bits of the signature~\cite{Simm84,Simm85,Des90,Des94,BDI+96,D88,BVS06}.

To ensure that the token uses proper signing randomness and to 
eliminate covert channels in the signatures,
we introduce \textit{firewalled ECDSA signatures} (Section~\ref{sec:crypto:san}).
To produce a firewalled signature, the token (holding a secret signing
key $\sk$ and message $m$) and browser (holding a public verification 
key $\pk$ and message $m$) run an interactive
protocol.
At the end of the protocol, the browser either (a) holds an ECDSA
signature on $m$ that is indistinguishable from an honestly
generated ECDSA signature using $\sk$ on $m$ or (b) outputs ``token failure.''
Crucially, our signing protocol is extremely efficient: it requires
the token to perform only two exponentiations in the group, 
while ECDSA requires one.

Finally, we make one privacy improvement to the standard U2F token design.
As described in Section~\ref{sec:bg}, to prevent cloning attacks, 
the U2F token includes a counter value in each authentication message it
sends to the relying party.
To use a minimal amount of non-volatile storage on the token,
today's tokens use a single global counter that increments after
each authentication attempt.
Using a global counter poses a privacy risk:
a coalition of relying parties can use this counter
value as a fingerprint to track a single user as she authenticates
to different services around the Web.

Using per-identity counters would protect against fingerprinting, but 
could require a large amount of storage space.
Because of hardware constraints we describe in Section~\ref{sec:count}, 
today's tokens use $2\times$ 2KB flash pages to store a single $23$-bit counter.
Using this strategy to implement 100 per-identity counters would require 400KB
of non-volatile storage---consuming almost all of our token's 512KB flash bank.

To allow for the use of distinct per-identity counters while respecting
the storage constraints of these weak tokens, we introduce a 
\textit{flash-optimized counter} data structure (Section~\ref{sec:count}).
Our design, which is inspired by log-structured file systems and
indexes~\cite{RO92,jffs,fawn,bufferhash,flashstore,silt,skimpystash,microhash}, 
can store 100 unique counters using only 6KB of flash,
while still supporting over $2^{22}$ total authentications. 

Putting these four pieces together yields the full \name design.

\subsection{Security goals}
\label{sec:goals:sec}
We now describe the security goals of the
\name design.

\paragraph{Protect against a malicious token.}
If the browser executes the protocol faithfully, 
the browser protects
against arbitrary misbehavior by the token.

To make this notion more formal, we adopt a definition inspired by the work
of Mironov and Stephens-Davidowitz on cryptographic reverse firewalls~\cite{MSD15}.
We say that the protocol protects against a malicious U2F token if an arbitrarily 
malicious relying party cannot distinguish the following two worlds:
\begin{itemize}[leftmargin=24pt]
  \item [W1] Server interacts with an ``ideal'' U2F token that executes
          the protocol faithfully.
  \item [W2] Server interacts with an arbitrarily malicious U2F token 
          that (a) sits behind an honest browser and (b) never causes the
          honest browser to abort.
\end{itemize}
A protocol that is secure under this definition prevents a malicious
token from exfiltrating secrets to a malicious relying party.

There are two nuances to this definition.
First, this security definition does not capture a number of critical 
potential covert channels, including timing channels and other 
other out-of-protocol attacks.
We discuss these attacks in Section~\ref{sec:goals:timing}.

Second, in our protocols, if the token sends a malformed message, the browser
may output ``Token Failure'' and refuse to continue processing messages.
We require that the relying party not be able to distinguish the two worlds described
above \textit{only for tokens that \textbf{never} cause an honest browser to abort} in this way.
We restrict our attention to tokens that 
do not cause the browser to abort for two reasons:
\begin{itemize}
  \item If the token fails in a detectable way,
        the user knows that something has gone wrong and can remediate
        by discarding the token and reinstalling the software on her machine.
        
  \item Furthermore, even if a malicious token triggers a failure
    event at a chosen time, it gains at most a negligible advantage in
    exfiltrating the user's cryptographic secrets to the outside
    world.  In Appendix~\ref{app:fail}, we explain why this is so.
\end{itemize}

\paragraph{Protect against a compromised browser.}
If the token executes the protocol faithfully, it should give the 
same level of protection against a malicious browser that U2F does today.
In particular, we consider an adversary that can:
\begin{itemize}
  \item passively observe the state of the browser at all times, and
  \item can take control of the browser at an adversary-chosen time~$T$.
\end{itemize}
We require that such an adversary cannot authenticate to any sites
that were registered with the token before time $T$,
except by interacting with the token.

Allowing the attacker to observe the state of the browser at all times models
the fact that it is notoriously difficult
to erase secrets from modern computers~\cite{czeskis2008defeating,garfinkel2004data,chow2005shredding}.
Because of this, if an attacker compromises a browser at time $T$, it can
likely recover bits of information about the browser's state from 
times $T' < T$.

Once the attacker takes control of the browser at time $T$, the
attacker can register for new sites without interacting with the honest token
(e.g., by forwarding U2F requests to an attacker-controlled token elsewhere).
So, the most we can require is that the token must participate 
when authenticating to sites registered at times $T' < T$.

\paragraph{Protect against token fingerprinting.}
If the token and browser execute the protocol faithfully, then
coalitions of malicious relying parties should not be able to ``fingerprint''
the token, with the aim of tracking the user across different
websites or linking different pseudonymous accounts on the same web site.

We define tracking-resistance using a ``real vs.~ideal'' security
game between a challenger and an adversary.
In the game, the challenger gives the attacker access to a \textit{registration} oracle
and an \textit{authentication} oracle.
These oracles implement the functionalities provided by the U2F token.
The attacker may invoke the registration oracle at most $I$ times
(e.g., $I \approx 100$), which
models an attacker that can coerce a user into registering its token 
under at most $I$ distinct identities at attacker-controlled websites.
The attacker may invoke the authentication oracle at most $A$ times 
(e.g., $A \approx 2^{22}$), which represents an upper bound on the 
number of authentication attempts the attacker can observe.

In the real world, the challenger implements these oracles using 
a \textit{single} U2F token.
In the ideal world, the challenger implements these oracles using
\textit{$I$ distinct} U2F tokens---one token per identity.
We say that the token protects against $I$-identity $A$-authentication 
fingerprinting if no efficient adversary can distinguish these two worlds 
with probability non-negligibly better than random guessing.

The browser UI should allow the user to distinguish 
between U2F registration and authentication, otherwise a single malicious
relying party could trick a user into registering their token more than
$I$ times, which could subvert \name's fingerprinting guarantees.
To prevent this attack, the browser could alert the user before the 
$I$th registration attempt, or the browser could prevent more than $I$
registrations with a single token.

When $I > 1$, this notion of tracking resistance is much stronger than today's
U2F tokens provide, since the use of a global authentication counter in 
today's U2F tokens provides a convenient device fingerprint.
We discuss this issue in detail in Section~\ref{sec:count}.

\subsection{Functionality goals}
\label{sec:goals:func}
To make any backdoor-resistant U2F design
usable in practice, it should be
backwards-compatible in the following ways.

\paragraph{No changes to the relying party (server).}
While a security-conscious user can upgrade their U2F token and browser to
use our new backdoor-resistant design, a security-conscious
user is \textit{not} able to upgrade the software running on their bank or
employer's web server. 
A security-conscious user should be able to protect herself against faulty or
malicious U2F tokens without waiting for all of the relying parties on the
Internet to upgrade to a new authentication protocol.

\paragraph{No changes to the U2F token hardware.}
While we might be able to achieve better performance by 
(for example) putting a faster processor in the token, by restricting ourselves to
today's hardware, we make sure that the deployment cost stays low.

\smallskip
\itpara{Using \name with many browsers.}
To audit the token across multiple browsers, we can piggyback
on existing mechanisms for syncing web bookmarks and preferences across 
a user's browser instances (e.g., Firefox~Sync).
The browser can store the token verification state---%
a total of a few kilobytes---alongside the data that the browser already syncs. 
In this way, all of the user's browser instances can effectively audit a single
U2F token.

The downside of this approach is that if an attacker manages to hijack any one
of a user's browser instances (or the sync server), the attacker can corrupt the
U2F verification state on all instances.
This trade-off seems somehow inherent; users who want the strongest protection
against browser and token compromise can always use a distinct U2F token on each
machine.

\subsection{Timing attacks and other covert channels}
\label{sec:goals:timing}

\name eliminates all ``in-protocol'' covert channels
that a malicious U2F token could use to exfiltrate information 
to a malicious relying party.
This still leaves open the possibility of ``out-of-protocol'' covert channels,
which we discuss here.

\sitpara{Timing.}
By modulating the amount of time it takes to respond to a relying party's authentication
request, a malicious token could leak information to the relying party without changing the
U2F protocol messages at all~\cite{fuzzytime}.
Essentially all exfiltration-prevention systems are vulnerable to some form
of timing attack~\cite{MSD15,cabuk2004ip}, and ours is no different.

To partially defend against this class of attacks, the browser could add a random delay
to U2F responses~\cite{fuzzytime,fuzzyfox}.
The browser could also delay the response so that it always takes a fixed amount
of time (e.g., five seconds) to complete.
Another defense would be to send a registration or authentication request
to the token only after the user presses a button on the
token---rather than waiting for a button press after the request is sent.
This modification would make it more difficult for a malicious relying party to
accurately measure the amount of time the token takes to complete a request.

\sitpara{Failure.}
By selectively refusing to complete the authentication protocol, the token can
leak roughly $\log_2 (T+1)$ bits of information to an outside relying party 
over the course of $T$ U2F interactions with the token.
(See Appendix~\ref{app:fail}.)
To mitigate the damage of this attack, the user of a token should discard the
token as faulty if it \textit{ever} fails to complete an authentication request. 

\sitpara{Physical attacks.}
If an adversary steals the token and can interact with it directly---without intermediation
by the browser---then a malicious token could leak all of its secrets to the thief
in response to a ``magic input.'' 
We do not attempt to protect against such physical attacks.

\sitpara{Other.} 
A backdoored U2F token could leak secrets outside of the flow of the
U2F protocol using a built-in Bluetooth or GSM modem. 
Or, a backdoored U2F token might be able to masquerade as a keyboard or other
input device, and coerce the browser or some other program on the user's
machine into leaking secrets to the outside world~\cite{TS18}.
Physical inspection of the token (e.g., to look for antennas) would detect the
most egregious physical backdoors.
Additional OS-level protections could defeat attacks via USB~\cite{Cinch,webusb,usbfilter}.

\section{Cryptographic building blocks}
\label{sec:crypto}

We first recall a number of standard cryptographic 
building blocks and then introduce
our new cryptographic protocols.

\sitpara{Notation.}
For an integer $q$, we use $\Z_q$ to denote the ring of
integers modulo $q$ and $\Z_q^*$ to denote its invertible elements.
For a variable $x$, $x \gets 3$ indicates assignment 
and $f(x) \deq x^2$ denotes definition.
For a finite set $\calS$, $r \getsr \calS$ denotes 
a uniform random draw from $\calS$.
We use ``$\bot$'' as a special symbol indicating failure.
When $\G = \langle g \rangle$ is a finite group, 
we always write the group notation multiplicatively.
So, an ECDSA public key has the form $X = g^x \in \G$.

An \textit{efficient} algorithm
is one that runs in probabilistic polynomial time.
When $A$ is a randomized algorithm, we use $A(x;r)$
to denote running $A$ on input $x$ with random coins~$r$.
A \textit{negligible} function is one whose inverse grows faster
than any fixed polynomial.
All of the cryptographic routines we define take a security parameter as an
implicit input and we require that they run in time polynomial in this
parameter.

\subsection{Standard primitives}
\label{sec:crypto:prim}

\paragraph{Digital signatures.}
A digital signature scheme over a message space $\calM$ 
consists of a triple of algorithms:
\begin{itemize}
  \item $\SigKeyGen() \to (\sk, \pk)$. Output
    a secret signing key $\sk$ and a public verification key $\pk$.
  \item $\SigSign(\sk, m) \to \sigma$. Output a signature
    $\sigma$ on the message $m \in \calM$ using the secret key~$\sk$.
  \item $\SigVerify(\pk, m, \sigma) \to \zo$. 
    Output ``$1$'' iff $\sigma$ is a valid signature on message $m$
    under public key $\pk$.
\end{itemize}
For all keypairs $(\sk, \pk)$ output by $\SigKeyGen$, 
for all messages $m \in \calM$, $\SigVerify(\pk, m, \SigSign(\sk, m)) = 1$.

We use the standard security notion for signature
schemes: \textit{existential unforgeability under chosen message attack}~\cite{GMR88}.
Informally, no efficient adversary should be able to construct a valid signature
on new a message of its choosing, even after seeing signatures on any number of other
messages of its choosing.

The current U2F standard mandates the use of ECDSA signatures~\cite{u2f},
summarized in Appendix~\ref{app:ecdsa},
so our protocols take the peculiarities of ECDSA into account.
That said, our protocols are compatible with more modern signatures schemes,
such as Schnorr-style~\cite{S91} elliptic-curve signatures~\cite{ed25519} and
BLS~\cite{BLS04}.

\paragraph{Verifiable random function (VRF).}
We use VRFs~\cite{VRF,DY05,IETF-VRF} defined over input space $\ID$ 
and output space $\Z^*_q$, for some prime integer $q$.
In our application, the input space $\ID$ is a set of ``identities''
(e.g., username-website pairs).
A VRF is a triple of algorithms:
\begin{itemize}
  \item $\VRFKeyGen() \to (\skVRF, \pkVRF)$. 
    Output a secret key and a public key.
  \item $\VRFEval(\skVRF, \id) \to (y, \pi)$.
    Take as input the secret key $\skVRF$, and
    an input $\id \in \ID$, and output a value $y \in \Z^*_q$ along with
    a proof $\pi$.
  \item $\VRFVerify(\pkVRF, \id, y, \pi) \to \zo$.
    Take as input the public key $\pkVRF$, 
    a purported input-output pair $(\id,y) \in \ID \times \Z^*_q$, and
    a proof $\pi$.
    Return ``$1$'' iff $\pi$ is a valid
    proof that $(\id,y)$ is an input-output pair.
\end{itemize}

To be useful, a VRF must satisfy the following standard notions, which
we state informally.
We refer the reader to prior work~\cite{NSEC5,NZ15} for formal definitions
of these properties. 
\begin{itemize}
  \item \textbf{Completeness.}
    The $\VRFVerify$ routine accepts as valid all proofs output by $\VRFEval$.
  \item \textbf{Soundness.}
    It is infeasible to find two
    input-output-proof triples $(\id, y, \pi)$ and $(\id, y',\pi')$ such that 
    (a) $y \neq y'$ 
    and (b) $\VRFVerify$ accepts both triples.
  \item \textbf{Pseudorandomness.}
    Informally, even if the adversary can ask for VRF output-proof pairs
    $(y_\id, \pi_\id)$ on $\id$s of its choosing,
    the adversary cannot distinguish a VRF output $y_{\id^*}$
    corresponding to an unqueried input $\id^*$ from a random point in $\Z^*_q$ with
    probability non-negligibly better than $1/2$.
\end{itemize}

We use a variant of the VRF construction of
Papadopoulos et al.~\cite{NSEC5},
which is secure in the random-oracle model~\cite{BR93} under
the Decision Diffie-Hellman assumption~\cite{B98}.
(We make only a trivial modification to their VRF construction to
allow it to output elements of $\Z^*_q$, rather than group elements.)

The only special property that we use of this VRF construction is that
its keypairs have the form $(x, g^x) \in \Z_q \times \G$, where $\G =
\langle g \rangle$ is a group of prime order~$q$, for $x$ random in
$\Z_q$.  
Since we otherwise make only black-box use of the VRF
construction, we do not discuss it further.

\subsection{New tool: Verifiable identity families (VIFs)}
\label{sec:crypto:vif}

Today's U2F tokens authenticate to every website under a distinct per-identity
public key.
Our protocol needs to ensure that the token samples its per-identity public
keys from the correct distribution 
(or, more precisely, a distribution indistinguishable from the correct one).
To do so, we introduce and construct a new primitive, 
called a \textit{verifiable identity family} (VIF).

The VIF key-generation routine outputs a master secret key and a master public key.
The master public key is essentially a commitment to a function from identities
$\id \in \ID$ (i.e., username-website pairs) to public keys for a signature scheme $\Sigma$.
Informally, the scheme provides the following functionality:
\begin{itemize}
\item 
Anyone holding the master secret key can produce the \textit{unique} 
public key $\pk_\id$ corresponding
to a particular identity $\id \in \ID$, and can also produce the corresponding
secret key $\sk_\id$ for the signature scheme $\Sigma$.

\item
The holder of the master secret key can prove to anyone
holding the master public key that $\pk_\id$ is really the unique public key
corresponding to the string $\id$.
\end{itemize}

Before giving a more precise definition, we first
explain how we use a VIF in our U2F token design. 
At the completion of the token
initialization process, the browser holds a VIF master public key, and the 
U2F token holds a VIF master secret key.
Whenever the user wants to register the token under a new
identity (e.g., $\id = \texttt{user@example.com}$), the browser sends the 
identity $\id$ to the token,
and the token returns a per-identity public key $\pk_\id$ 
to the browser, 
along with a proof
that the public key was computed correctly.
The browser verifies the proof to convince itself that 
$\pk_\id$ is the 
correct public key for $\id = \texttt{user@example.com}$ 
under the token's master public key.
The soundness property of the VIF ensures that there is a \textit{unique} public key 
that the browser will accept from the token for this site. 
In this way, we can prevent a malicious token from leaking bits of information
to a malicious website via a non-random or otherwise malformed public key.

Furthermore, we require the VIF-generated public keys to 
be pseudorandom: a malicious browser should not be able to predict
what the VIF-generated public key will be for an identity $\id$ 
without querying the token for $\pk_\id$.
This prevents a malicious browser from registering the token at new
websites without the token's participation.

Finally, we require the VIF to satisfy an unforgeability property: 
a malicious browser should not be able to forge signatures that verify
under VIF-generated public keys.
If the VIF satisfies unforgeability, a malicious browser cannot learn
useful information about the token's secrets, even after it watches
the token authenticate to many websites.

\textit{Hierarchical wallets}~\cite{bip32}, used
in Bitcoin, are closely related to VIFs.
The key difference is that hierarchical wallet schemes either
(a) do not allow the holder of a public key $\pk_\id$ to verify
that it is the unique key corresponding to a master public key $\mpk$, or
(b) do not satisfy $\Sigma$-pseudorandomness, defined below.
Both properties are crucial for our application.

\paragraph{Syntax and definitions.}
A VIF is a triple of efficient algorithms, 
defined with respect to a signature scheme $\Sigma = (\SigKeyGen, \SigSign, \SigVerify)$:
\begin{itemize}
  \item $\VIFKeyGen() \to (\msk, \mpk)$. 
        Output a master secret key $\msk$ and a master public key $\mpk$.
  \item $\VIFEval(\msk, \id) \to (\sk_\id, \pk_\id, \pi)$.
        Given an identity string $\id \in \ID$, output the keypair
        $(\sk_\id, \pk_\id)$ corresponding to that identity, as well
        as a proof $\pi$ that $\pk_\id$ is well formed.
        The keypair $(\sk_\id, \pk_\id)$ must be a valid keypair for the 
        signature scheme $\Sigma$.
  \item $\VIFVerify(\mpk, \id, \pk_\id, \pi) \to \zo$.
        Verify that $\pk_\id$ is the public key for identity $\id \in \ID$
        corresponding to master public key $\mpk$ using the proof $\pi$.
        Return ``$1$'' iff the proof is valid.
\end{itemize}
The last two algorithms are deterministic.

We require the following four security properties to hold.
The first two properties are almost identical to the properties required from a VRF. 
The last two properties are new, so we define them formally in Appendix~\ref{app:sig:vifdefs}.
\begin{itemize}
\item \textbf{Completeness.}
  For any $(\msk, \mpk)$ output by $\VIFKeyGen$ and any identity 
  $\id \in \ID$, if $(\sk_\id, \pk_\id, \pi_\id) \gets \VIFEval(\msk, \id)$
  then $\VIFVerify(\mpk, \id, \pk_\id, \pi_\id) = 1$.

\item \textbf{Soundness.}  
      For all efficient adversaries $\calA$, if we sample
      $(\msk, \mpk) \gets \VIFKeyGen()$ and then run 
      $\calA(\msk, \mpk)$, the probability that $\calA$ outputs two 
      identity-key-proof triples $(\id, \pk_\id, \pi)$ 
      and $(\id, \pk'_\id, \pi')$ such that 
      $\pk_\id \neq \pk'_\id$ and
      $\VIFVerify(\mpk, \id, \pk_\id, \pi) = 1$ 
      and $\VIFVerify(\mpk, \id, \pk'_\id, \pi') = 1$,
      is negligible in the (implicit) security parameter.

\item \textbf{$\Sigma$-Pseudorandomness.}   
      Even if the adversary can see many $\Sigma$-type signature public keys
      $\{\pk_{\id_1}, \pk_{\id_2}, \dots \}$ for identities of its choosing,
      and even if the adversary can see $\Sigma$-type signatures on messages
      of its choosing using the corresponding secret 
      keys $\{\sk_{\id_1}, \sk_{\id_2}, \dots \}$,
      the adversary still cannot distinguish the true public key $\pk_{\id^*}$
      for some identity $\id^* \not \in \{\id_1, \id_2, \dots \}$
      from a fresh public key $\pk_{\textsf{rand}}$ output by $\SigKeyGen()$
      with probability non-negligibly more than $1/2$.

      This holds even if the adversary may ask for signatures under
      the secret key corresponding to the challenge public key
      (which is either $\pk_{\id^*}$ or $\pk_\textsf{rand}$).

\item \textbf{$\Sigma$-Unforgeability.} 
      The keypairs output by $\VIFEval(\msk,\cdot)$ are
      safe to use as $\Sigma$-type signing keypairs. 
      In particular, even if the adversary can see many $\Sigma$-type 
      public keys output by $\VIFEval(\msk,\cdot)$
      for identities of its choosing, and even if the adversary can see
      $\Sigma$-type signatures on messages $\{m_1, m_2, \dots \}$
      of its choosing using the corresponding secret keys, 
      the adversary still cannot produce a signature forgery,
      namely a triple $(\id^*, m^*, \sigma^*)$ such that
      (a) the adversary never asked for a signature on $m^*$ under identity $\id^*$ and
      (b) $\sigma^*$ is a valid signature on message $m^*$ under the public key 
      $\pk_{\id^*}$ that $\VIFEval(\msk, \id^*)$ outputs.

\end{itemize}

\begin{figure}
{\small
\begin{framed}
\textbf{Our VIF construction for ECDSA.}
The ECDSA signature scheme uses a group $\G$ of prime order~$q$.
An ECDSA keypair is a pair $(y, g^y) \in \Z_q \times \G$, for $y \getsr \Z_q$.
The construction makes use of a VRF $\calV = (\VRFKeyGen, \VRFEval, \VRFVerify)$ 
that maps $\ID$ into~$\Z^*_q$ and uses keypairs of the form $(x, g^x) \in \Z_q
\times \G$.

We instantiate the three VIF routines as follows:
\begin{itemize}
  \item $\VIFKeyGen() \to (\msk, \mpk)$.
    \begin{itemize}
      \item Choose a random $x \getsr \Z_q$.
      \item Run $(\skVRF, \pkVRF) \gets \VRFKeyGen()$.
      \item Output $\msk = (x, \skVRF),\ \ \mpk = (g^x, \pkVRF)$.
    \end{itemize}

  \item $\VIFEval(\msk, \id) \to (\sk_\id, \pk_\id, \pi)$. 
    \begin{itemize}
      \item Parse the master secret key $\msk$ as a pair $(x, \skVRF)$.
      \item Run $(y, \piVRF) \gets \VRFEval(\skVRF, \id)$.
      \item Set $(\sk_{\id},\pk_\id) \gets (x y, g^{xy}) \in \Z_q \times \G$
      and $\pi \gets (y, \piVRF)$.
      \item Output $(\sk_\id, \pk_\id, \pi)$.
    \end{itemize}

  \item $\VIFVerify(\mpk, \id, \pk_\id, \pi) \to \zo$
    \begin{itemize}
      \item Parse the master public key $\mpk$ as a pair $(X, \pkVRF)$, where $X \in \G$.
            Parse the public key $\pk_\id$ as a group element $Y\in \G$.
            Parse the proof $\pi$ as a pair $(y, \piVRF)$, where $y \in \Z^*_q$.
      \item Output ``$1$'' iff 
            (a) $Y = X^y \in \G$ and 
            (b) $\VRFVerify(\pkVRF,\allowbreak \id, y, \piVRF) = 1$.
    \end{itemize}
\end{itemize}
\end{framed}
}
\caption{Our verifiable identity family construction for ECDSA.}
\label{fig:vifconstr}
\end{figure}

\paragraph{Our construction.}
Our construction of a VIF for the ECDSA signature scheme
appears as Figure~\ref{fig:vifconstr}.
To sketch the idea behind the construction: the master public key 
consists of a standard ECDSA public key $X = g^x$ and a VRF public key.
To generate the public key for an identity $\id \in \ID$, the VIF evaluator 
applies a VRF to $\id$ to get a scalar $y \in \Z_q$. 
The public key for identity $\id$ is then $X^y \in \G$.

The completeness and soundness of this VIF construction 
follow immediately from the completeness and soundness of the
underlying VRF. 
We prove the following theorem in Appendix~\ref{app:sig:proofs}, 
as Theorems~\ref{thm:vifpr} and~\ref{thm:vifuf}:
\begin{theorem}[Informal]\label{thm:vif-main}
The VIF construction of Section~\ref{sec:crypto:vif}
satisfies $\Sigma$-pseudorandomness and $\Sigma$-unforgeability 
in the random-oracle model, when $\Sigma$
is the Idealized ECDSA signature scheme of Appendix~\ref{app:ecdsa},
instantiated (1) with a secure VRF and (2) in a group in which
the discrete-log problem is infeasible.
\end{theorem}
Our VIF construction is also secure when used with the Schnorr digital signature
scheme~\cite{S91}.
The proof of security follows a very similar structure to that in Appendix~\ref{app:sig:proofs}.

Proving that our VIF construction 
satisfies $\Sigma$-unforgeability when $\Sigma$ is the ECDSA
signatures scheme is not entirely trivial.
The reason is that when using our VIF construction to generate secret keys for 
a sequence of identities $\id_1, \id_2, \dots, \id_n$, the corresponding secret
keys are \textit{related}.
That is, the secret ECDSA signing keys are of the form:
$(\sk_{\id_1}, \sk_{\id_2}, \dots, \sk_{\id_n}) = (x y_1, x y_2, \dots, x y_n)$,
such that the attacker knows $y_1, y_2, \dots, y_n \in \Z_q$. 
To prove that our VIF scheme satisfies $\Sigma$-unforgeability, we must prove that
using ECDSA to sign messages with related keys in this way does not, for example,
allow the attacker to recover the master secret key $x$.
Morita et al.~\cite{RKA15} show attacks 
on Schnorr and DSA-style signatures schemes when attacker may \textit{choose}
the relations $y_1, \dots, y_n \in \Z_q$.
In our case, the $y$-values are sampled using a VRF, so their attacks
do not apply. 

\subsection{New tool: Fast collaborative key generation}\label{sec:crypto:keygen}
Corrigan-Gibbs et al.~present a protocol that allows a network router
and certificate authority (CA) to collaboratively generate an ECDSA
keypair~\cite{randkeys}.
By directly applying their result, we get 
a protocol in which the browser and token can jointly
generate an ECDSA keypair in such a way that (1) the browser learns nothing
about the secret key and (2) the token cannot bias the public key that the
protocol outputs. 

We present a protocol that achieves the same security properties as theirs, 
but that requires the token to compute only a single exponentiation in the group $\G$,
while theirs requires at least three. 
Since our protocol runs on a computationally limited U2F token,
this factor of three yields a non-trivial speedup.

The master keypair for our VIF construction of Section~\ref{sec:crypto:vif}
consists of two ECDSA-style keypairs, so the browser and token 
can use this protocol to jointly generate the VIF master keypair.

The protocol is parameterized by a group $\G$ of prime order~$q$.
At the end of the protocol, 
the browser outputs an ECDSA public key $X \in \G$ 
and the token either (a) outputs an ECDSA secret key $x \in \Z_q$
such that $X = g^x$, or (b) outputs the failure symbol~``$\bot$.''
We use the standard notion of 
\textit{statistical closeness} of probability distributions~\cite{goldreich}.
If two distributions are statistically close, no efficient adversary
can distinguish them.

The protocol maintains the following properties:
\begin{itemize}
  \item \textbf{Completeness.}
        If both parties are honest, the token never outputs 
        the failure symbol ``$\bot$.''
  \item \textbf{Bias-free.}
        At the conclusion of a protocol run, the following 
        two properties hold: 
        \begin{itemize}
          \item If the browser is honest, it holds a public key drawn
                from a distribution that is statistically close to
                uniform over $\G$.
          \item If the token is honest, it holds a private key drawn
                from a distribution that is statistically close to uniform
                over $\Z_q$, provided that the browser \textit{never} causes the token 
                to output~``$\bot$.'' 
        \end{itemize}
  \item \textbf{Zero knowledge (Malicious browser learns nothing).}
        If the token is honest, the browser learns nothing from
        participating in the protocol except the output public key.
        Formally, there exists an efficient simulator that can simulate
        a malicious browser's view of the protocol given only 
        the public key as input, provided that the browser never 
        causes the token to output ``$\bot$.''

        With more work, we can give a relaxed definition for browsers that 
        cause the token to output $\bot$ with noticeable probability.
        The extension is straightforward, so we use the simpler definition. 
\end{itemize}

\paragraph{Protocol.}
The protocol can use any non-interactive statistically hiding commitment scheme~\cite{DPP97,NOVY98},
but for simplicity, we describe it using a hash function $H: \Z_q \times \Z_q \to \Z_q$ 
that we model as a random oracle~\cite{BR93}.
\begin{enumerate}
  \item The browser commits to a random value $v \in \Z_q$.
        That is, the browser samples $v \getsr \Z_q$ and $r \getsr \Z_q$ and 
        sends $c \gets H(v,r)$ to the token.
  \item The token samples $v' \getsr \Z_q$ and sends $V' \gets g^{v'} \in \G$
        to the browser.
  \item The browser opens its commitment to $v$ by sending
        $(v,r)$ to the token. The browser 
        outputs $X \gets V' \cdot g^v \in \G$ as the public key.
  \item The token checks that $H(v,r) = c$.
        If so, the token accepts and outputs $x \gets v + v' \in \Z_q$.
        Otherwise, the token outputs~``$\bot$.''
\end{enumerate}

Completeness follows immediately.
In Appendix~\ref{app:keygen}, we prove:
\begin{theorem}[Informal]
The key-generation protocol above is bias-resistant
and zero knowledge, when we model $H$ as a random oracle.
\end{theorem}

\subsection{New tool: Firewalled ECDSA signatures}
\label{sec:crypto:san}

During authentication, the relying party sends a challenge string
to the U2F token, and the token signs it using ECDSA. 
We want to ensure that the token cannot exfiltrate any secrets by embedding
them in the bits of the signature.
If the U2F standard used a \textit{unique} signature scheme,
such as RSA Full Domain Hash or BLS~\cite{BLS04}, this would be a non-issue.
Since ECDSA is not unique, the token could use its
signatures to leak its secrets. 

To eliminate exfiltration attacks, we introduce \textit{firewalled signatures},
inspired by cryptographic reverse firewalls~\cite{MSD15}.
A firewalled signature scheme is a standard digital signature scheme $\Sigma$, along
with an interactive signing protocol that takes place between a signer $\calS$ 
and a firewall $\calF$. 
In a run of the signing protocol, the signer takes as input a secret signing key $\sk$
for $\Sigma$ and a message~$m$.
The firewall takes as input the public key $\pk$ corresponding to~$\sk$ and the same
message $m$.

At the conclusion of the protocol, the firewall holds a signature~$\sigma$ on $m$
that validates under the public key $\pk$, and the firewall learns nothing else.
Furthermore, $\sigma$ is indistinguishable from an ``ideal'' $\Sigma$-type
signature on the message $m$ using the secret key~$\sk$.
In this way, the signer cannot exfiltrate
any bits of information in $\sigma$ itself.

\paragraph{Definition and syntax.}
A firewalled signature scheme consists of a 
standard digital signature scheme $\Sigma = (\SigKeyGen, \SigSign, \SigVerify)$ 
over a message space $\calM$, along with an interactive protocol
between a signer $\calS$ and a firewall $\calF$.
For bitstrings $x, y \in \zo^*$, let $[\calS(x) \leftrightarrow \calF(y)]$
denote the string that $\calF(y)$ outputs when interacting with $\calS(x)$.
Let $\View_{\calF}[\calS(x) \leftrightarrow \calF(y)]$ denote the transcript
of messages that $\calF$ observes in this interaction, along with $\calF$'s input.

Then the signing protocol $(\calS, \calF)$ must satisfy:
\begin{itemize}
  \item \textbf{Correctness.}
        For all $(\sk, \pk)$ output by $\SigKeyGen$, 
        for all messages $m \in \calM$,
        if $\sigma \gets [\calS(\sk, m) \leftrightarrow \calF(\pk, m)]$, then
        $\SigVerify(\pk, \sigma, m) = 1$.

  \item \textbf{Exfiltration resistance.}
        Informally, as long as a malicious signer $\calS^*$ never causes 
        the honest firewall $\calF$ to reject, then
        no efficient algorithm can tell whether a signature $\sigma$ was generated
        using the standard signing algorithm or via the interactive protocol 
        between $\calS^*$ and $\calF$.

        Let $\calS^*$ be an efficient algorithm such that 
        for all messages $m \in \calM$, 
        for all choices of the random coins of $\calS^*$ and $\calF$, 
        we have that $[\calS^*(\sk, m) \leftrightarrow \calF(\pk, m)] \neq \bot$.
        Then the following distributions are computationally indistinguishable
        \begin{align*}
          \Dreal &= \big\{ \sigma \gets [\calS^*(\sk, m) \leftrightarrow \calF(\pk, m)] \big\} \\
          \Dideal &= \big\{ \sigma \gets \SigSign(\sk, m) \big\},
        \end{align*}
        where we sample $(\sk, \pk) \gets \SigKeyGen()$.

  \item \textbf{Zero knowledge.}
        Informally, the only thing that a malicious firewall can learn from
        participating in the protocol is a valid signature on the message~$m$.

        Let $\calF^*$ be an efficient firewall that never
        causes the honest signer $\calS$ to output ``$\bot$.''
        (As in Section~\ref{sec:crypto:keygen}, handling the more general case 
        of arbitrary efficient $\calF^*$ is straightforward.)
        Then, there exists an efficient algorithm $\Sim$ such that
        for all such $\calF^*$ and and all messages $m \in \calM$, the
        following distributions are computationally indistinguishable:
        \begin{align*}
          \Dreal &= \big\{ \View_{\calF^*}[\calS(\sk, m) \leftrightarrow \calF^*(\pk, m)] \big\}\\
          \Dideal &= \big\{ \Sim \big(\pk,\, m,\, \SigSign(\sk,m) \big) \big\},
        \end{align*}
        where we sample $(\sk, \pk) \gets \SigKeyGen()$. 

\end{itemize}

\begin{figure}
\begin{framed}
  {\small
\textbf{Signing protocol for firewalled ECDSA.}
For a keypair $(\sk, \pk) \gets \EKeyGen()$, and a message $m \in \calM$
in the ECDSA message space $\calM$,
the protocol takes place between a signer $\calS(\sk, m)$
and a firewall $\calF(\pk, m)$:
\begin{enumerate}
  \item The parties $\calS$ and $\calF$ engage in the key-generation
        protocol of Section~\ref{sec:crypto:keygen}.
        The signer plays the role of the token and the firewall
        plays the role of the browser.
        If the protocol aborts, $\calF$ outputs $\bot$. 
        Otherwise, at the end of the protocol, the signer $\calS$ holds a value
        $r \in \Z_q$ and the firewall $\calF$ holds value $R = g^r \in \G$.\label{proto:san:1}
  \item The signer $\calS$ executes $\sigma \gets \ESign(\sk, m; r)$, 
        using $r \in \Z_q$ as the signing nonce, and
        sends the signature $\sigma$ to~$\calF$.
    \item The firewall uses Equation~(\ref{eq:ecdsa}) of Appendix~\ref{app:ecdsa}
        to compute the value $\wR \in\{g^{r'}, g^{-r'}\}$,
        where $r'$ is the signing nonce used to generate $\sigma$.
        The firewall ensures that $\EVerify(\pk, m, \sigma) = 1$ and $R \in \{\wR, 1/(\wR) \}$, and
        outputs $\bot$ if either check fails.
  \item As described in Appendix~\ref{app:ecdsa}, given a valid ECDSA signature $\sigma$
        on a message $m$ under public key $\pk$, anyone can produce a second valid signature
        $\bar \sigma$.
        The firewall $\calF$ outputs a random signature
        from the set $\{\sigma, \bar \sigma\}$.
\end{enumerate}
}
\end{framed}
\caption{Our protocol for generating firewalled ECDSA signatures.}
\label{proto:san}
\end{figure}

\paragraph{Construction.}
It is possible to construct firewalled signatures for
any signature scheme using pseudo-random functions and
zero-knowledge proofs.
We develop a much more efficient 
special-purpose signing protocol that enables us to firewall 
plain ECDSA signatures without resorting to general zero-knowledge techniques.

The ECDSA signature scheme is a tuple of algorithms
$(\EKeyGen, \ESign, \EVerify)$, whose definition we recall in 
in Appendix~\ref{app:ecdsa}.
Recall that ECDSA signatures use a fixed group $\G = \langle g \rangle$
of prime order~$q$.

There are two key ideas behind our construction:
\begin{itemize}
  \item First, the only randomness in an ECDSA signature 
        comes from the signer's random choice of a ``signing nonce'' 
        ${r \getsr \Z_q}$.
        (In fact, $r$ is sampled from $\Z^*_q = \Z_q \setminus \{0\}$, but
        this distinction is meaningless in practice, since the two distributions
        are statistically indistinguishable.)
  \item Second, given an ECDSA public key $\pk$, a message $m$, and a valid ECDSA 
        signature $\sigma$ on $m$ for $\pk$, anyone can recover a value 
        $\wR$, such that, if $r \in \Z_q$ is the signing nonce,
        then either $\wR = g^{r}$ or $\wR = g^{-r}$.
        (See Equation~(\ref{eq:ecdsa}) in Appendix~\ref{app:ecdsa}.)
\end{itemize}

In our signing protocol, 
the signer and firewall can use the collaborative key-generation protocol of 
Section~\ref{sec:crypto:keygen} to generate this $(r, g^r)$ pair in a way
that ensures that $r$ is distributed uniformly at random over $\Z_q$.
(The firewall actually only recovers 
$\wR = \cramped{g^{\pm r}}$, but we can patch this with one extra step.)
The signer then uses the random value $r$ to produce the ECDSA signature.
Finally, the firewall checks that the signature is valid and
that the signer used the correct nonce $r$ to generate it, 
and then the firewall outputs a rerandomized version of the signature.
Figure~\ref{proto:san} describes the full protocol.

We prove the following theorem in Appendix~\ref{app:san}:
\begin{theorem}[Informal]\label{thm:san}
If the key-generation protocol 
of Section~\ref{sec:crypto:keygen} is bias free and 
zero knowledge, then 
the protocol of Figure~\ref{proto:san} is exfiltration resistant
and zero knowledge.
\end{theorem}

\section{Storing many counters in little space}
\label{sec:count}

As described in Section~\ref{sec:bg}, the token must send a counter value to
the relying party that increases with each authentication. 

The simplest implementation, and the one used on Yubico's U2F token
(and originally on our token as well),
keeps a global counter that increments with each
authentication attempt. 
However, using a single global counter---rather than a counter per
identity---entails both security and privacy risks:
\begin{enumerate}
  \item \textit{Failure to always detect clones.}
    Say that an attacker clones a token and logs into site $S_1$ then $S_2$.
    Then say that the token owner (with the true token) logs into site $S_2$ then $S_1$.
    If the global counter had value $c$ when the token was cloned, 
    $S_1$ and $S_2$ will each see a counter value of $c+1$ followed by $c+2$, 
    and neither site will detect that the token was cloned.
  \item \textit{Token fingerprinting.}
    A single global counter value serves as a token fingerprint that can allow 
    colluding sites to track a user across sites. 
    Consider a token that authenticates at colluding sites $S_1$, $S_2$, and $S_1$ again.
    If these sites see the counter values $(c, c+1, c+2)$, 
    for some value $c$, during these three authentication attempts, the sites can make
    an educated guess that the same user is authenticating at these two sites.
\end{enumerate}

\subsection{Goals and hardware constraints}
\label{sec:count:hw}

We need a data structure with an 
operation $\Init() \to \st$ that initializes a fresh counter state,
and an operation 
$\Inc(\st, \id)\allowbreak \to (\st_\ms{new}, c_\id)$, that increments the counter value
associated with string $\id \in \zo^*$ and returns the new state $\st_\ms{new}$
and counter value~$c_\id$.

The functionality goals for our counter scheme
are parameterized by a maximum number of identities $I$ used
and a maximum number of increments $A$ (corresponding to $A$ authentications).
Informally, we would like the following properties:
\begin{itemize}
  \item \textbf{No worse than a global counter.}
        Counter values always increase.
        Also, after performing $t$ increments from a fresh state with any sequence of $\id$s, the counter value returned (for any $\id$) 
        is never larger than $t$.
  \item \textbf{As good as $I$ independent counters.}
        As long as the user authenticates to at most $I$ sites, 
        the structure behaves as if each identity $\id$ is assigned 
        an independent counter.
\end{itemize}

\paragraph{Flash hardware constraints.}
Any counter scheme we design must conform to the constraints 
of our token's flash hardware.
Typical low-cost U2F tokens have similar requirements.

Our token uses two 256KB banks of NOR flash, divided into 2KB pages
of 32-bit words.
The flash controller supports random-access reads, word-aligned writes,
and page-level erases. 
In flash, an erase sets all of the bits in the page to \ttO,
and subsequent writes can only shift bits to \ttZ or leave them unchanged. 
An interrupted write or erase operation leaves the modified bits
in an indeterminate state.

Because of the physics of the flash, the hardware allows
\begin{itemize}
  \item only \textbf{erasing an entire page} at a time---it is not possible to 
        erase only a single word,
  \item at most \textbf{eight writes to each 32-bit word} between erases, and
  \item at most \textbf{50,000 program/erase cycles per page} over the lifetime
          of the token, and
    \item any number of reads between erases.
\end{itemize}
Furthermore, the counter implementation must be robust to interrupts---if 
the user yanks the U2F token out of their machine during authentication, 
the counter should still end up in a coherent state and the counter value
presented to any site should be greater than the previous value.
This prevents a user from being locked out of their account.

\subsection{A ``straw-man'' counter scheme}
These implementation constraints make it surprisingly expensive, in terms of
flash space, to store even a small counter.

Using a single flash page, we can implement a counter that counts up to 4,096.
To do so, we encode the counter value in \textit{unary}: the counter
begins in the erased state (all \ttO{}s).
To increment the counter, we zero out (``knock down'') the first four non-zero bits
in the page.
In this way, we respect the constraint of at most eight writes per 32-bit word 
and we never have to erase the page until the counter reaches its maximum value.
There are 512 words per page, so we can count up to a maximum value of
$512\cdot (32/4) = 2^{12}$.

Putting two such counters together---one for the low-order bits and one
for the high-order bits---yields a two-page counter that counts up to $2^{24}$,
or roughly 16 million.
With a little more work, we can make this design robust to power failures
and other interrupts, allowing us to count up to $2^{23}$.
Popular U2F tokens today use
a very similar two-page counter design. 

As discussed above, one way to provide better cloning protection and
unlinkability is to use per-identity counters.
If we wanted to have 100 per-identity counters on the token, this would
require \textit{\textbf{two hundred}} flash pages, 
or over 400KB of flash storage---%
on a token with only 512KB of storage total.
In contrast, our design stores 100 counters that can support a
total of over $2^{22}$ increment operations using
only \textit{\textbf{three}} flash pages.

\subsection{New tool: Flash-friendly counters}

In \name, we implement the authentication counters 
using a logging data structure~\cite{RO92}. 
The structure conceptually consists of a fixed-length table of identity 
count pairs $\langle \id, c_\id \rangle$, along with a variable-length 
list of identities (the ''log'').
To increment the counter associated
with identity $\id$, we append $\id$ to the log. 
When the log grows large, we run 
a ``garbage collection'' operation: we use the log to update the 
table of identity-count pairs, and then erase the log.

\medskip

\begin{figure}
\centering
\usetikzlibrary{decorations.pathreplacing}
\usetikzlibrary{patterns}

\begin{tikzpicture}

\newcommand{\logtop}{3}
\newcommand{\logright}{2}
\newcommand{\colwidth}{0.3}
\newcommand{\barc}{yellow!20!white}
\newcommand{\oldc}{black!30!white}
\newcommand{\cntc}{green!20!white}
\newcommand{\labeloffset}{-0.7}

\newcommand{\barsize}{0.5}
\fill [fill=blue!05!white] (0,\logtop-4*\barsize) rectangle (2*\colwidth,\logtop);
\fill [fill=\barc] (2*\colwidth,\logtop-\barsize) rectangle (2,\logtop);
\fill [fill=\barc] (2*\colwidth,\logtop-2*\barsize) rectangle (2,\logtop-\barsize);
\fill [fill=\barc] (2*\colwidth,\logtop-3*\barsize) rectangle (1,\logtop-2*\barsize);
\fill [fill=\barc] (2*\colwidth,\logtop-4*\barsize) rectangle (2,\logtop-3*\barsize);

\draw [thick,draw=black!30!white,pattern color=black!30!white,pattern=north east lines] (0,0) rectangle (2,\logtop-4*\barsize);
\draw [thick] (0,\logtop-\barsize) rectangle (2,\logtop);
\draw [thick] (0,\logtop-2*\barsize) rectangle (2,\logtop-\barsize);
\draw [thick] (0,\logtop-3*\barsize) rectangle (1,\logtop-2*\barsize);
\draw [thick] (0,\logtop-4*\barsize) rectangle (2,\logtop-3*\barsize);

\draw[thick,dotted] (\colwidth,0+2*\barsize) -- (\colwidth,\logtop);
\draw[thick,dotted] (2*\colwidth,0+2*\barsize) -- (2*\colwidth,\logtop);

\draw [decorate,decoration={brace,amplitude=3pt,mirror,raise=2pt},yshift=0pt]
(2*\colwidth,0) -- (\logright,0) node [midway,yshift=-8pt] {\scriptsize 126-bit identifier};
\node at (1,\labeloffset) {\small \textbf{Log page}}; 
\node[text width=1in,align=center] at (1.3,\logtop+0.2) {\scriptsize Data}; 
\node[text width=0.1in,align=left,rotate=90] at (0.45,\logtop+0.2) {\scriptsize Type}; 
\node[text width=0.1in,align=left,rotate=90] at (0.15,\logtop+0.2) {\scriptsize Invalid?}; 

\node[circle,fill=blue,inner sep=0.75pt,color=blue] at (0.8,\logtop-2.5*\barsize) {};
\draw[->,thick,color=blue] (0.8,\logtop-2.5*\barsize) to[out=25,in=-190] (3,1.25);

\newcommand{\writeZero}[2]{\node at (#1,#2) {\scriptsize \texttt{0}};}
\newcommand{\writeOne}[2]{\node at (#1,#2) {\scriptsize \texttt{1}};}
\newcommand{\writeOneg}[2]{\node[color=\oldc] at (#1,#2) {\scriptsize \texttt{1}};}
\newcommand{\writeDots}[2]{\node[color=\oldc] at (#1,#2) {\scriptsize \texttt{$\dots$}};}
\newcommand{\writeText}[3]{\node at (#1,#2) {\scriptsize #3};}
\newcommand{\writeTextg}[3]{\node[color=\oldc] at (#1,#2) {\scriptsize #3};}

\writeZero{0.5*\colwidth}{\logtop-0.5*\barsize};
\writeZero{0.5*\colwidth}{\logtop-1.5*\barsize};
\writeZero{0.5*\colwidth}{\logtop-2.5*\barsize};
\writeZero{0.5*\colwidth}{\logtop-3.5*\barsize};
\writeOneg{0.5*\colwidth}{\logtop-4.5*\barsize};
\writeOneg{0.5*\colwidth}{\logtop-5.5*\barsize};

\writeZero{1.5*\colwidth}{\logtop-0.5*\barsize};
\writeZero{1.5*\colwidth}{\logtop-1.5*\barsize};
\writeOne{1.5*\colwidth}{\logtop-2.5*\barsize};
\writeZero{1.5*\colwidth}{\logtop-3.5*\barsize};
\writeOneg{1.5*\colwidth}{\logtop-4.5*\barsize};
\writeOneg{1.5*\colwidth}{\logtop-5.5*\barsize};
\writeOneg{2.5*\colwidth}{\logtop-4.5*\barsize};
\writeOneg{2.5*\colwidth}{\logtop-5.5*\barsize};
\writeOneg{3.5*\colwidth}{\logtop-4.5*\barsize};
\writeOneg{3.5*\colwidth}{\logtop-5.5*\barsize};
\writeOneg{4.5*\colwidth}{\logtop-4.5*\barsize};
\writeOneg{4.5*\colwidth}{\logtop-5.5*\barsize};

\writeDots{5.7*\colwidth}{\logtop-4.5*\barsize};
\writeDots{5.7*\colwidth}{\logtop-5.5*\barsize};

\writeText{\colwidth + 0.5*\logright}{\logtop-0.5*\barsize}{$\ms{Hash}(\id_6)$};
\writeText{\colwidth + 0.5*\logright}{\logtop-1.5*\barsize}{$\ms{Hash}(\id_{8})$};
\writeText{\colwidth + 0.5*\logright}{\logtop-3.5*\barsize}{$\ms{Hash}(\id_{11})$};

\draw [decorate,decoration={brace,amplitude=3pt,raise=2pt},yshift=0pt]
  (0,0) -- (0,2*\barsize) node [midway,rotate=90,yshift=8pt] {\scriptsize Log space};

\draw [thick,draw=black!30!white,pattern color=black!30!white,pattern=north east lines] (3,0) rectangle (5,\logtop);
\foreach \i in {2,3,4} {
\writeOneg{3+0.5*\colwidth+\i*\colwidth}{\logtop-0.5*\barsize};
}
\writeDots{3+5.75*\colwidth}{\logtop-0.5*\barsize};
\fill [fill=blue!05!white] (3,\logtop-\barsize) rectangle (3+1.5*\colwidth,\logtop);
\fill[fill=\barc] (3,\logtop-2*\barsize) rectangle (4.4,\logtop-\barsize);
\fill[fill=\barc] (3,\logtop-3*\barsize) rectangle (4.4,\logtop-2*\barsize);
\fill[fill=\barc] (3,\logtop-4*\barsize) rectangle (4.4,\logtop-3*\barsize);
\fill[fill=\barc] (3,\logtop-5*\barsize) rectangle (4.4,\logtop-4*\barsize);

\fill[fill=\cntc] (4.4,\logtop-2*\barsize) rectangle (5,\logtop-\barsize);
\fill[fill=\cntc] (4.4,\logtop-3*\barsize) rectangle (5,\logtop-2*\barsize);
\fill[fill=\cntc] (4.4,\logtop-4*\barsize) rectangle (5,\logtop-3*\barsize);
\fill[fill=\cntc] (4.4,\logtop-5*\barsize) rectangle (5,\logtop-4*\barsize);
\fill[fill=\cntc] (4.4,\logtop-6*\barsize) rectangle (5,\logtop-5*\barsize);

\fill[fill=red!30!white] (3,\logtop-6*\barsize) rectangle (4.4,\logtop-5*\barsize);
\draw[thick] (3,\logtop-\barsize) -- (3+2*\colwidth,\logtop-\barsize);

\draw [thick] (3,0) rectangle (5,\logtop - \barsize);
\draw [thick] (3,\logtop - \barsize) rectangle (3 + 1.5*\colwidth,\logtop);
\node at (4,\labeloffset) {\small \textbf{Active data page}}; 
\draw[thick,dotted] (3 + 1.4,0) -- (3+1.4,\logtop-\barsize);
\draw[thick] (3,\logtop-2*\barsize) -- (3+\logright,\logtop-2*\barsize);
\draw[thick] (3,\logtop-3*\barsize) -- (3+\logright,\logtop-3*\barsize);
\draw[thick] (3,\logtop-4*\barsize) -- (3+\logright,\logtop-4*\barsize);
\draw[thick] (3,\logtop-5*\barsize) -- (3+\logright,\logtop-5*\barsize);
\node at (3+0.75*\colwidth,\logtop-0.5*\barsize) {\texttt{4}};


\node[anchor=east] at (3.05,\logtop-0.15) {\parbox{0.5in}{\flushright \scriptsize Serial\\[-2pt]number\\[-2pt](16 bits)}}; 

\writeText{3 + 0.5*\logright - 0.3}{\logtop-1.5*\barsize}{$\ms{Hash}(\id_1)$};
\writeText{3 + 0.5*\logright - 0.3}{\logtop-2.5*\barsize}{$\ms{Hash}(\id_3)$};
\writeText{3 + 0.5*\logright - 0.3}{\logtop-3.5*\barsize}{$\ms{Hash}(\id_4)$};
\writeText{3 + 0.5*\logright - 0.3}{\logtop-4.5*\barsize}{$\ms{Hash}(\id_5)$};
\writeText{3 + 0.5*\logright - 0.3}{\logtop-5.5*\barsize}{\texttt{<overflow>}};

\writeText{4.7}{\logtop-1.5*\barsize}{\texttt{37}};
\writeText{4.7}{\logtop-2.5*\barsize}{\texttt{41}};
\writeText{4.7}{\logtop-3.5*\barsize}{\texttt{15}};
\writeText{4.7}{\logtop-4.5*\barsize}{\texttt{2}};
\writeText{4.7}{\logtop-5.5*\barsize}{\texttt{29}};

\draw [decorate,decoration={brace,amplitude=3pt,mirror,raise=2pt},yshift=0pt]
  (4.4,0) -- (5,0) node [midway,yshift=-8pt] {\scriptsize 34-bit counter};

\node at (7,\labeloffset) {\small \textbf{Inactive data page}}; 
\foreach \i in {2,3,4} {
\writeOneg{6+0.5*\colwidth+\i*\colwidth}{\logtop-0.5*\barsize};
}
\writeDots{6+5.75*\colwidth}{\logtop-0.5*\barsize};
\node[anchor=east] at (6.05,\logtop-0.15) {\parbox{0.5in}{\flushright \scriptsize Serial\\[-2pt]number\\[-2pt](16 bits)}}; 

\writeTextg{6 + 0.5*\logright - 0.3}{\logtop-1.5*\barsize}{$\ms{Hash}(\id_1)$};
\writeTextg{6 + 0.5*\logright - 0.3}{\logtop-2.5*\barsize}{$\ms{Hash}(\id_2)$};
\writeTextg{6 + 0.5*\logright - 0.3}{\logtop-3.5*\barsize}{$\ms{Hash}(\id_4)$};
\writeTextg{6 + 0.5*\logright - 0.3}{\logtop-4.5*\barsize}{$\ms{Hash}(\id_5)$};
\writeTextg{6 + 0.5*\logright - 0.3}{\logtop-5.5*\barsize}{\texttt{<overflow>}};
\draw[thick,dotted,color=\oldc] (6 + 1.4,0) -- (6+1.4,\logtop-\barsize);

\writeTextg{7.7}{\logtop-1.5*\barsize}{\texttt{35}};
\writeTextg{7.7}{\logtop-2.5*\barsize}{\texttt{4}};
\writeTextg{7.7}{\logtop-3.5*\barsize}{\texttt{10}};
\writeTextg{7.7}{\logtop-4.5*\barsize}{\texttt{2}};
\writeTextg{7.7}{\logtop-5.5*\barsize}{\texttt{26}};

\draw [thick,draw=\oldc,pattern color=\oldc,pattern=north east lines] (6,0) rectangle (8,\logtop);
\draw [thick,draw=\oldc] (6,0) rectangle (8,\logtop - \barsize);
\writeText{3 + 0.5*\logright - 0.3}{\logtop-5.5*\barsize}{\texttt{<overflow>}};

\writeText{4.7}{\logtop-1.5*\barsize}{\texttt{37}};
\writeText{4.7}{\logtop-2.5*\barsize}{\texttt{41}};
\writeText{4.7}{\logtop-3.5*\barsize}{\texttt{15}};
\writeText{4.7}{\logtop-4.5*\barsize}{\texttt{2}};
\writeText{4.7}{\logtop-5.5*\barsize}{\texttt{29}};

\draw[color=\oldc] (6,\logtop-2*\barsize) -- (6+\logright,\logtop-2*\barsize);
\draw[color=\oldc] (6,\logtop-3*\barsize) -- (6+\logright,\logtop-3*\barsize);
\draw[color=\oldc] (6,\logtop-4*\barsize) -- (6+\logright,\logtop-4*\barsize);
\draw[color=\oldc] (6,\logtop-5*\barsize) -- (6+\logright,\logtop-5*\barsize);

\fill [fill=blue!05!white] (6,\logtop-\barsize) rectangle (6+1.5*\colwidth,\logtop);
\node at (6+0.75*\colwidth,\logtop-0.5*\barsize) {\texttt{3}};
\draw [thick] (6,\logtop-\barsize) rectangle (6+1.5*\colwidth,\logtop);

\end{tikzpicture}
\caption{Counter data structure. 
  The log stores 
  hashes of identifiers (type $\ttZ$) and
  pointers to entries in the active data page (type $\ttO$).}
  \label{fig:counter}
\end{figure}
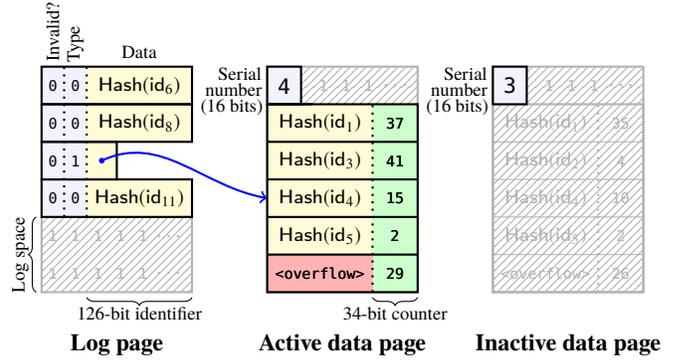

Our implementation uses three flash pages---a log page and two data pages---%
and it can store up to $I=100$ distinct counters 
and supports $A = 6.4$ million authentications.
We give a graphical depiction of the structure in Figure~\ref{fig:counter}.

Each data page stores a table of $I$ identity-count pairs, 
along with an ``overflow'' counter $c_\overflow$ and a page serial number.
The data page with the larger serial number is the \textit{active}
page, and the other data page is the \textit{inactive} page.
The log page stores the log.
As a space-saving optimization, we actually only store hashes of identities,
using a hash output large enough to avoid collisions in sets of at most $I$ identities.

The counter value associated with identity $\id$ is the sum of 
(a) the number of times $\id$ appears in the log and 
(b) the counter value $c_\id$ associated with $\id$ 
    in the active data page (if one exists) 
    or the value $c_\overflow$ (if none exists).

To increment the counter associated with identity $\id$, we
can simply append the the string ``$\id$'' to the end of the log page.
To save space, when incrementing the counter
associated with an identity $\id$ that already
appears in the active data page, we write a pointer into the log that
references the location of $\id$ in the active page.
In the common case, when the user repeatedly authenticates under the same few
identities, this optimization allows us to fit $8\times$ more identities in the
log.

The only complexity arises when the log page is nearly full.
When this happens, we ``garbage collect'' the log:
We pick $I$ identities, first from the most recently used identities in the log page,
and any remaining from the identities associated with the largest counter values in
the active data page. 
We write these identity-count pairs into the inactive data page along with
the new value of the overflow counter $c_\overflow$.
We compute this value as the maximum of the old overflow value 
and the counts associated with any $\id$s
that we ``evicted'' from the active data page during garbage collection.
We then increment the serial number on the inactive data page, which 
makes it the new active page, and we erase the log page.

We order these operations carefully to ensure that even if the token loses
power during these operations, it can later recover the correct counter state. 
To detect failures during log updates,
each log entry is encoded with an ``invalid'' bit.
When the token writes the last word of the log entry, it clears this bit
to mark the entry as valid.
To detect failures during garbage collection, the token
writes a special symbol to the log before garbage collection
with the serial number of the active page.

To compute the maximum number of increments: we erase each page at most
once during garbage collection.
In the worst case, where every authentication increments a unique counter value,
the maximum number of increments is 50,000 (the maximum
number of erases) times 128 (the number of hashes that can fit in the log),
which yields 6.4 million increments total.
In the best case, where we only increment at most $I=100$ counters,
the maximum number of increments is 128 (the number of hashes we add to the log
before garbage collecting the first time) plus 49,999 (the maximum number of
erases minus one) times 1,024 (the number of pointers that can fit in the log),
for a total of 51 million increments.
To increase this maximum value, a manufacturer could 
use multiple log pages.

\section{The complete \name protocols}

\begin{figure}
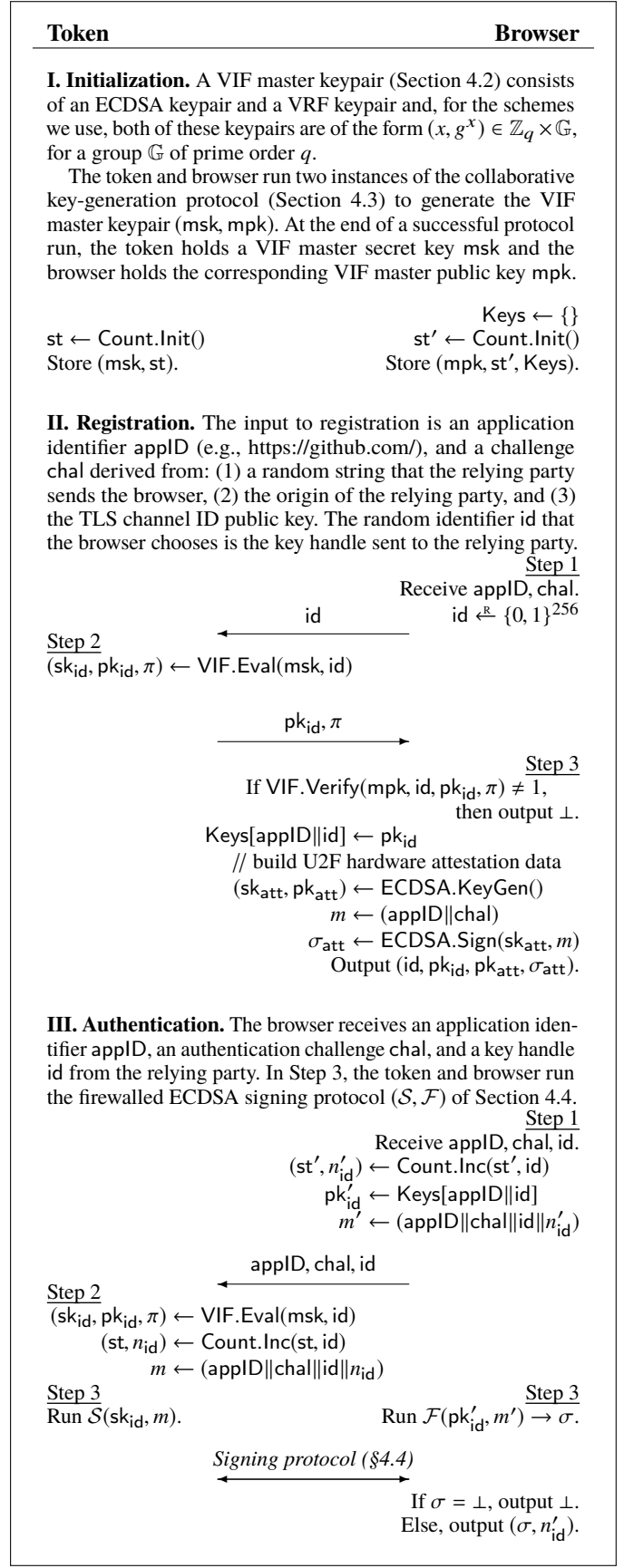

\centering
  \begin{framed}
  \centering
\newcounter{protoCounter}

  \newcommand{\ProtoStepN}[1]{\underline{\smash{Step #1}}}
\newcommand{\ProtoStep}{\refstepcounter{protoCounter}%
  \ProtoStepN{\arabic{protoCounter}}}

\begingroup
\addtolength{\jot}{-0.2em}
{\small
\begin{tabular*}{\columnwidth}{l @{\extracolsep{\fill}} c r}
\textbf{\normalsize Token}& & \textbf{\normalsize Browser}\\
\hline\\
\multicolumn{3}{l}{\parbox{0.94\columnwidth}{\textbf{I. Initialization.} 
A VIF master keypair (Section~\ref{sec:crypto:vif}) consists
of an ECDSA keypair and a VRF keypair and, for the schemes we use,
both of these keypairs are of the form $(x, g^x) \in \Z_q \times \G$, for a group $\G$
of prime order~$q$.

\quad The token and browser run two instances of the collaborative
    key-generation protocol
(Section~\ref{sec:crypto:keygen}) to generate the VIF master keypair
$(\msk, \mpk)$.
At the end of a successful protocol run, the token holds a
VIF master secret key $\msk$ and the browser holds the corresponding
  VIF master public key $\mpk$.}\vspace{1em}}\\[32pt]
  && $\keys \gets \{\}$\\
$\stD \gets \Init()$&& $\stB \gets \Init()$\\
  Store $(\msk, \stD)$. && Store $(\mpk, \stB, \keys)$.\\[16pt]
\multicolumn{3}{l}{\parbox{0.94\columnwidth}{\textbf{II. Registration.} 
The input to registration is an application identifier 
$\appID$ (e.g., https://github.com/),
and a challenge $\chal$ derived from:
  (1) a random string that the relying party sends the browser, 
  (2) the origin of the relying party, and 
  (3) the TLS channel ID public key.
The random identifier $\id$ that the browser chooses is
the key handle sent to the relying party.}}  \\[16pt]
&&\ProtoStep\\
&& Receive $\appID, \chal$.\\
&& $\id \getsr \zo^{256}$\\[-18pt]
\figureLeftArrow{$\id$}\\[6pt]
\ProtoStep\\
\multicolumn{3}{l}{$(\sk_\id, \pk_\id, \pi) \gets \VIFEval(\msk, \id)$}\\[6pt]
\figureRightArrow{$\pk_\id, \pi$}\\[12pt]
  &&\ProtoStep\\
\multicolumn{3}{r}{If $\VIFVerify(\mpk, \id, \pk_\id, \pi) \neq 1$,~~~~~~\ }\\
  \multicolumn{3}{r}{then output $\bot$.}\\
\multicolumn{3}{r}{
  $\begin{aligned}
    \keys[\appID \| \id] &\gets \pk_\id\\
    /\!/\ \text{build U2F}\ & \text{hardware attestation data} \\
    (\sk_\batch, \pk_\batch) &\gets \EKeyGen()\\
    m&\gets (\appID \| \chal)\\
    \sigma_\batch &\gets \ESign(\sk_\batch, m)
  \end{aligned}$}\\
  \multicolumn{3}{r}{Output $(\id, \pk_\id, \pk_\batch, \sigma_\batch)$.}\\[16pt] 
\multicolumn{3}{l}{\parbox{0.94\columnwidth}{\textbf{III. Authentication.}
The browser receives an application identifier $\appID$, 
an authentication challenge $\chal$, and a key handle $\id$
from the relying party.
In Step~\ref{step:sign}, the token and browser run the
firewalled ECDSA signing protocol $(\calS, \calF)$ of Section~\ref{sec:crypto:san}.
  \setcounter{protoCounter}{0}}}  \\[12pt]
&&\ProtoStep\\
&& Receive $\appID, \chal, \id$.\\
  \multicolumn{3}{r}{$\begin{aligned}
    (\stB, n'_\id) &\gets \Inc(\stB, \id)\\
    \pk'_\id &\gets \keys[\appID \| \id]\\
    m' &\gets (\appID \| \chal \| \id \| n'_\id)
  \end{aligned}$
  }\\[2pt]
\figureLeftArrow{$\appID, \chal, \id$}\\[6pt]
\ProtoStep\\
\multicolumn{3}{l}{
  $\begin{aligned}
    (\sk_\id, \pk_\id, \pi) &\gets \VIFEval(\msk, \id)\\
    (\stD, n_\id) &\gets \Inc(\stD, \id)\\
    m &\gets (\appID \| \chal \| \id \| n_\id)
  \end{aligned}$
}\\[6pt]
\ProtoStep \label{step:sign}&& \ProtoStepN{\arabic{protoCounter}}\\
Run $\calS(\sk_\id, m).$ && Run $\calF(\pk'_\id, m') \to \sigma$.\\
\figureBothArrow{\textit{Signing protocol (\S\ref{sec:crypto:san})}}\\[12pt]
&& If $\sigma = \bot$, output $\bot$.\\
&& Else, output $(\sigma, n'_\id)$.
\end{tabular*}
}
\endgroup
\end{framed}

  \vspace{-1.3ex}
  \caption{%
An overview of the \name protocols.
For clarity, the figure does not describe the 
optimizations of Section~\ref{sec:impl:opt}, 
and omits the U2F control fields
(``control byte'' and ``user-presence byte'').}
\label{fig:proto}
\end{figure}

We show how to assemble the cryptosystems of
Section~\ref{sec:crypto} with the counter design
of Section~\ref{sec:count} into the full \name system
(Figure~\ref{fig:proto}). 

\paragraph{I. Token initialization.}
\label{sec:proto:init}
\name adds an initialization step to the U2F protocol.
In this step, the \name token and browser collaboratively 
generate a master keypair $(\msk, \mpk)$ for our verifiable
identity family construction from Section~\ref{sec:crypto:vif}.
Since a VIF keypair for our construction consists of 
two ECDSA-style keypairs (one for signing and one for the VRF),
the token and browser can generate both keypairs using
our key-generation 
protocol of Section~\ref{sec:crypto:keygen}.
When using our protocol to generate the master VIF keypair:
\begin{itemize}
  \item an honest browser is assured it is using a master public
        key sampled from the correct distribution and
\item an honest token is assured that the browser learns nothing about
  its master secret key (other than what can be inferred from the master public key).
\end{itemize}

\sitpara{Alternative: Load keys from an external source.}
A user could generate the master VIF keypair $(\msk, \mpk)$ 
on a single machine---separate from both the token and browser---%
and could load $\msk$ onto the \name token and $\mpk$ onto the browser.
If the user puts $\msk$ in offline storage, she can use it to recover
all of her on-token secrets if she loses the physical token.

\medskip

The browser and token also both initialize a counter data structure
(Section~\ref{sec:count}) to store the U2F authentication counters.
If both parties are honest, then the browser's counter state will replicate the
state on the token.

\paragraph{II. Registration with website.}
During registration, the relying party sends the browser an application
identity (e.g., service name) and a random challenge, as shown
in Figure~\ref{fig:u2f}. 
The token must return
(1) a ``key handle'' that the relying party will pass back to the token
    during authentication,
(2) a public key,
(3) an attestation certificate containing the attestation public key, and
(4) a signature over the registration message with the attestation secret key
    corresponding to the attestation public key.
The last two fields are used for U2F's hardware attestation feature.
For privacy reasons, the Chrome browser by default generates a fresh
random self-signed attestation certificate on every registration and uses
this to sign the registration message~\cite{chrome-batch}.
\name follows Chrome's design.

In \name, the browser chooses a random key handle $\id \getsr \zo^{256}$ 
during registration. 
Using a random key handle ensures that if the user registers twice at
the same website, the user receives two independent identities.
The browser then sends $\id$ to the token and the token 
uses the VIF and its master secret key 
to derive the unique public key $\pk_\id$ corresponding to this identity.
The token returns this key along with a proof of correctness $\pi$ to the browser.
The browser then checks the proof using the VIF master public key.

\com{
For privacy reasons, \name does not use
hardware attestation signing keys (see Section~\ref{sec:bg}).
Instead, the browser generates a fresh attestation keypair on each 
registration attempt and it signs the registration challenge using this keypair.
}

\paragraph{III. Authentication with website.}
During authentication, the relying party sends the browser
(1) an application identifier, 
(2) a challenge, 
and (3) a key handle,
as shown in Figure~\ref{fig:u2f}.
The token must return a counter value and a signature over the
authentication request and counter value.

In \name, the browser sends the application identifier, challenge, and key
handle to the token.
The token then uses the VIF to derive the per-identity signing key $\sk_\id$ for
this identity and the browser can recover the per-identity public key $\pk_\id$ it
received from the token during registration.
The browser and token can both use their respective counter state
to derive the counter value for identity (i.e., key handle) $\id$.
Using the application identifier, challenge, and counter value, the browser
and token can both derive the message $m$ to be signed.

The browser and token then engage in the firewalled 
ECDSA signing protocol of Section~\ref{sec:crypto:san}
to jointly produce a signature $\sigma$ 
on the challenge message $m$.
If the signing protocol succeeds, the 
browser returns the signature $\sigma$ and counter
value $n'_\id$ to the relying party.

One detail is that the message that the token signs
includes a ``user-presence byte'' that indicates whether
the user physically touched the token during authentication.
The U2F v1.2 specification~\cite{u2f} has a mode in which
the user need not touch the token during authentication.
To prevent data leakage via the user-presence byte,
the browser must ensure that it is set to \texttt{0x01}
if the relying party required user presence and \texttt{0x00}
otherwise.

\paragraph{Security analysis.}
We argue informally
that our protocol provides the three security properties defined in
Section~\ref{sec:goals:sec}.
We discussed timing and selective failure attacks in
Section~\ref{sec:goals:timing}. 

We prove each claimed property from Section~\ref{sec:goals:sec}
using a hybrid argument.
We begin with the real interaction between adversary and challenger,
as defined in that section.
We then make a number of incremental modifications to this interaction,
arguing at each step that the adversary cannot distinguish the 
modified interaction from the real one.
Finally, we reach an ``ideal'' interaction, in which the adversary
cannot break security, which completes the proof.
 
\itpara{Property 1: \name protects against a malicious token.} \\
Proof idea:
The proof proceeds in four steps. 
We first replace the master public key
$\mpk$ obtained by the browser with one generated by $\VIFKeyGen()$.
The bias-freeness of the key-generation protocol
permits this change.

Then, we replace the per-identity public keys $\pk_\id$ sent
by the token during registration with ones generated by $\VIFEval()$.
By the soundness of the VIF, this change modifies the adversary's
advantage by at most a negligible amount.
By VIF pseudorandomness, we can then replace 
the per-identity public keys by fresh public 
keys output by $\EKeyGen()$.

Finally, we replace the signatures produced during authentication
with signatures generated using $\ESign()$.
The exfiltration resistance property of the firewalled 
signature scheme ensures that this change increases
the adversary's success probability by at most a negligible amount.
At this point, all the values sent to the relying party are sampled
from the ideal distribution, as required.

\pagebreak[3]
\itpara{Property 2: \name protects against a compromised browser.} \\
Proof idea:
To prove the claim, we first replace the master secret key
$\msk$ given to the honest token with one generated by $\VIFKeyGen()$.
The bias-freeness of the key-generation protocol
permits this change.

The zero knowledge property of the firewalled signature scheme
implies that we can replace the signing protocol between the honest
token and corrupt browser run during authentication
(Step~III.\ref{step:sign} in Figure~\ref{fig:proto}) by having the
honest token send a properly generated signature $\sigma$ to the
browser.  The remaining protocol transcript elements can be generated
by the zero-knowledge simulator.

Then, protection against a compromised browser follows
immediately from the VIF's unforgeability property.
That is, to authenticate to a previously registered website without interacting
with the token, the corrupt browser must produce a valid signature on a
challenge, under $\pk_\id$, for some identity $\id$,
without querying the token for a signature on this challenge.
(Here, we rely on the fact that before the browser is compromised,
it samples the identities $\id$ from a space large enough to avoid
collisions.)  

\itpara{Property 3: \name protects against token fingerprinting.} \\
Proof idea;
First, we replace the adversarial token with an ideal token.
By the ``Protects against malicious token'' property (above), this
change cannot improve the adversary's success probability by
a non-negligible amount.

Next, we replace our counter construction with $I=100$ independent counters. 
The properties of our counter design ensure that this change does
not increase the adversary's advantage, as long as the adversary
performs no more than $A=6.4$ million authentications.

Finally, we replace---one at a time---the per-identity public keys $\pk_\id$
generated by the VIF with independent random public keys.
With each replacement, the pseudorandomness of the VIF ensures that the
relying party cannot notice the change.
At this point, the adversary's view in the two worlds is identical, so it
has no hope of distinguishing them.

\section{Implementation}
\label{sec:impl}

The source code for our implementation is
available at \url{https://github.com/edauterman/u2f-ref-code}
and \url{https://github.com/edauterman/true2f}.

\subsection{System architecture}
Our \name implementation consists of: 
(i) an extension to the Chrome browser, (ii) a local agent that
runs alongside the browser, and (iii) a hardware token. 

\sitpara{Browser Extension:} 
Our browser extension forwards U2F requests to a local agent process that runs
\name with the token, and is compatible with relying parties that support U2F.
We built on an existing extension for U2F development~\cite{u2frefcode}.
To do so, we wrote roughly 50 lines of JavaScript code.
Our extension does not yet sync token state
via a user's browser profile (Section~\ref{sec:goals:func}). 

\sitpara{Local Agent:} The local agent runs the \name protocol with the token.
We built the local agent on top of an existing U2F reference 
implementation~\cite{u2frefcode} using the OpenSSL library.
To do so, we wrote roughly
3,800 lines of C/C++ code: 
2,000 for running the cryptographic protocol, 
700 for the simulating the token's flash-friendly counter (to enforce
correct counter values), 
800 for parsing and encoding, and
300 for the interface to the extension. 

\sitpara{Token:} We implemented the \name protocol
in firmware on a standard U2F hardware token
used by Google employees. 
It uses an ARM SC-300 processor 
clocked at 24~MHz and has a cryptographic accelerator 
with an interface exposed for certain operations (see Table~\ref{tab:crypto-ops}).
Our implementation on the token required
adding or modifying roughly 2,000 lines of C code: 1,400 for the protocol,
and 600 for the counter.

\subsection{Cryptographic optimizations}
\label{sec:impl:opt}

We have implemented a number of optimizations to minimize the cost
of the \name protocols, and we sketch these here.

\paragraph{Browser-assisted hash-to-point.}
To evaluate the VRF used in our VIF construction, the token must compute a hash
function $H_\G$ that maps bitstrings into points on the P256 elliptic curve.
We implement $H_\G$ using the ``try-and-increment'' 
method~\cite[Section 3.3]{BLS04}.
If P256 is defined over the field $\F_p$, the try-and-increment method
requires the token to compute the square root modulo $p$ of the \textit{first}
quadratic residue (i.e., square modulo $p$) 
in a sequence of values $Z = z_1, z_2, z_3, \dots \in \F_p$.
As Table~\ref{tab:crypto-ops} shows,
computing square roots on the token is costly---%
more than $5\times$ slower than computing signatures.
(There are other methods for implementing $H_\G$, but these either do not apply
to the P256 curve or also require computing square roots in
$\F_p$~\cite{I09,B10,Elligator,Elligator2}.)

The token can outsource most of the $H_\G$ computation to the browser. 
To do so, we take advantage of the fact that when $\F_p$ is the P256 base
field, if $z \in \F_p$ is a quadratic non-residue, then $-z \in \F_p$ is a
quadratic residue.
The outsourcing then works as follows:
the token and browser both compute the sequence $Z$.
Let $z_\ell$ be the first quadratic residue in $Z$.
Then all values $z_1, \dots, z_{\ell-1}$ are quadratic non-residues.
For each such value $z_i$, the browser sends the token the square root 
$r_i$ of $-z_i \in \F_p$.
The token checks that $\cramped{r_i^2} = -z_i \in \F_p$, 
for all $i \in \{1, \dots, \ell-1\}$.
Finally, the browser sends the square root $r_\ell$ of $z_\ell$
and the token checks that $\cramped{r_\ell^2} = z_\ell \in \F_p$.
Since \textit{checking} a square-root computation 
(i.e., computing $\cramped{x^2} \bmod p$)
is $\approx 256\times$ faster that \textit{computing} a square root modulo $p$, 
this outsourcing is profitable for the token.

This optimization brings the cost for the hash-to-point operation
down to 2.5ms, from an unoptimized cost of 199ms.

\begin{table}[t]
\centering
  \newcolumntype{Y}{>{\centering\arraybackslash\cellcolor{yellow!25}}b{1em}}
\newcolumntype{Z}{>{\raggedleft\arraybackslash}b{1em}}
\newcolumntype{Q}{>{\centering\arraybackslash}b{1em}}
\newcommand{\yy}{\cellcolor{yellow!30}}
{\centering\arraybackslash}
\renewcommand{\arraystretch}{1.1}
{\small \setlength\tabcolsep{3pt}
  \begin{tabular}{lcr@{\hskip12pt}ZZZZQQ}
    \multicolumn{3}{b{2in}}{
\caption{Cost of various operations
    on the token, averaged over 100 runs,
    and the expected number of each operation required
    per authentication attempt.
  ``HW?'' indicates use of the token's crypto
    accelerator.  \vspace{-18pt}
    }\label{tab:crypto-ops}}&\rot{No opts.}
  &\rot{+ Fast keygen}
  &\rot{+ VRF caching}
  &\rot{+ Hash assist}
  &\yrot{\textbf{\name} (+ all)}
  &\rot{U2F}\\
  \textbf{Operation} & \textbf{HW?} & \textbf{Time} (\textmu s) & \multicolumn{6}{c}{\textbf{Ops.~per auth.}}\\
  \midrule
  SHA256 (128 bytes) & Y & 19 & 5 & 5 & 3 & 5 & \yy 3 & 1\\
  $x + y \in \Z_q$ & N & 36 & 17 & 16 & 2 & 15 & \yy 1 & 0\\
  $x \cdot y \in \Z_q$ & N & 409 & 9 & 8 & 2 & 11 & \yy 1 & 0\\ 
$g^x \in \G$ & Y & 17,400 & 7 & 5 & 3 & 7 & \yy 1 & 0\\
  $\ESign$ & Y & 18,600 & 1 & 1 & 1 & 1 & \yy 1 & 1\\
  $g \cdot h \in \G$& N & 25,636 & 1 & 0 & 1 & 1 & \yy 0 & 0\\
  $\sqrt x \in \Z_q$ & N & 105,488 & 2 & 2 & 0 & 0 & \yy 0 & 0\\
\bottomrule
\end{tabular}
}

\end{table}

\paragraph{Caching VRF outputs.}
When registering or authenticating with a site with identity $\id$, 
the token uses our VIF construction (Section~\ref{sec:crypto:vif})
to compute the per-identity signing key $\sk_\id$.
To compute $\sk_\id$, the token must evaluate the VRF at the point $\id$
to get a VRF output $y_\id \in \Z_q$.

Even with the hash-to-point optimization described above, 
evaluating the VRF is relatively costly, 
since it requires computing an exponentiation in $\G$. 
We eliminate the need for the costly VRF computation during the
authentication phase entirely by computing a message authentication code
(MAC)~\cite{BCK96} and using the browser as an off-token cache.
During initialization, the token generates and stores a MAC secret key.
The MAC is only relevant when the token is honest, 
so the token can generate this MAC key on its own.

During registration, we have the token first compute the VRF
output $y_\id$.
Then, the token computes a MAC tag $\tau_\id$ over the tuple
$\langle \id, y_\id \rangle$ and sends the triple $\langle \id, y_\id, \tau_\id \rangle$
to the browser, along with its registration response message.
The MAC tag never leaves the browser, so the token cannot
use it to exfiltrate data.

Later on, when the browser wants the token to authenticate to site $\id$, 
the browser sends the token the triple $\langle \id, y_\id, \tau_\id \rangle$
along with its authentication request.
After verifying the MAC tag, 
the token can use $y_\id$ to generate
the per-identity secret key $\sk_\id$ without having
to recompute the VRF at the point~$\id$.

Since computing and verifying a MAC just requires a few invocations of SHA256,
this optimization brings the cost of evaluating the VIF down to $0.47$ms,
compared with 73.75ms when using only the hash-to-point optimization.

\section{Evaluation}

\begin{figure}[t]
\centering
\input{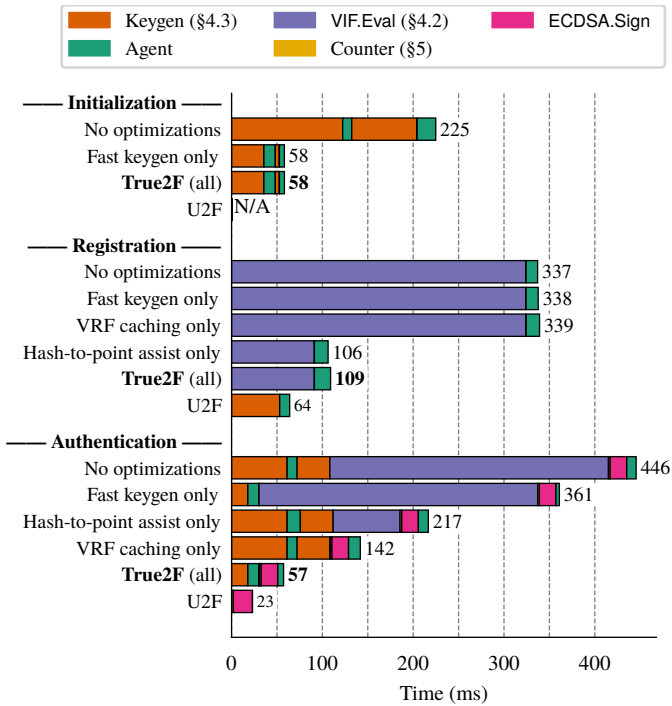}
\caption{Protocol execution time for \name, U2F, and less optimized
  variants of \name, averaged across 100 runs.
  For all measurements, we instrument the token to not wait for user touch.
  The standard deviation for all measurements is less than 2ms.
  }
\label{time-breakdown}
\end{figure}

We evaluate \name on the hardware token described in Section~\ref{sec:impl}
and an agent running on an Intel Xeon W-2135 processor at 3.8GHz.
Our \name implementation uses 85KB bytes of token flash space total:
75KB of code and 10KB for keys and counters.
For comparison, a plain U2F implementation uses
70KB of space: 64KB of code and 6KB of keys and counters.

\paragraph{Protocol execution time.} 
Figure~\ref{time-breakdown} gives the time to execute the
initialization, registration, and authentication protocols for 
\name, U2F, and partially optimized variants of \name, timed from when the agent
begins the protocol with the token to when the agent is ready to respond to the
browser extension. 

A \name registration (109ms) is $1.7\times$ slower than a U2F registration (64ms).
Figure~\ref{time-breakdown} demonstrates the effect of the hash-to-point 
optimization (Section~\ref{sec:impl:opt}) during registration---this optimization cuts execution
time by over 230ms.
A \name authentication takes \NameAuthMs{}ms, which is $2.3\times$
slower than an unprotected U2F authentication (\UAuthMs ms).
The cost of $g^x$ computations dominates the authentication cost:
our protocol requires four (two on the token and two on the agent)
while unprotected U2F requires only one, to generate the ECDSA signature.
Our VRF caching optimization (Section~\ref{sec:impl:opt}) essentially
eliminates the cost of generating the per-identity signing key using
$\VIFEval$.

\begin{figure}[t]
\centering
\input{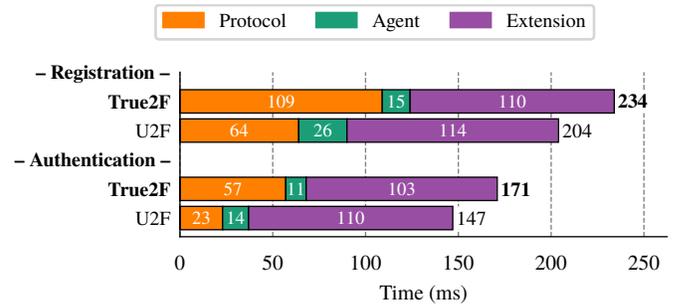}
    \caption{Total time to complete a FIDO U2F request.}
\label{end-to-end}
\end{figure}

\paragraph{End-to-end latency.} 
Figure~\ref{end-to-end} shows the total time for registering 
or authenticating with \name and unprotected U2F, measured from 
the time that the webpage invokes the Javascript API to
the time the API call returns.
As Figure~\ref{end-to-end} shows, the total time for \name authentication is 171ms,
compared with 147ms for unprotected U2F, which amounts to a 16\% slowdown.

\begin{wrapfigure}{r}{0.5\columnwidth}
\input{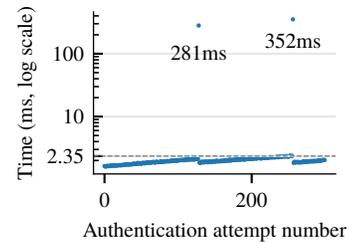}
\caption{Counter update time
  with synchronous garbage collection.}
\label{counter}
\end{wrapfigure}

\paragraph{Counters.}
Figure~\ref{counter} shows the time required to increment a
counter when using our log-structured counter design. 
When there is no need to garbage collect the log, an
increment takes between 1.63ms and 2.35ms. 
The difference in increment time is a function of the log size: to look up the
counter value associated with a relying party, the token must read through the entire log. 
To make increments constant time, a hardened implementation would
read through the entire log flash page---not just to the end of the log.
To avoid triggering an expensive garbage-collection operation during 
registration or authentication, we can 
execute garbage collection asynchronously 
(e.g., while the token is plugged in to a USB port but is idle).

The simple global counter design requires 0.43ms to increment, compared with 
a worst-case time of 2.35ms for our design.
In absolute terms, the costly cryptographic operations we execute in \name 
mask this 2ms counter-update time.

\paragraph{Browser sync state.}
To support using the same token with many browsers, 
\name requires syncing state across multiple browser instances with
the same user profile. 
If a user registers the token with $I$ identities, an upper bound
on the size of the sync state is $4162+97 \cdot I$ bytes.
For the fixed state, $4$KB is needed to store the counters, and 66 bytes
are needed for the VIF master public key. Each additional
identity requires syncing a 32-byte MAC, a 32-byte VRF output, and a 33-byte
public key.
With 100 registered identities, this state amounts to less than $14$KB,
which is far less than the 100KB of state that the Chrome browser will 
sync per extension~\cite{chrome-storage}.

\section{Related work}
\label{sec:rel}

The threat of hardware backdoors (``hardware Trojans'')
motivates our work~\cite{TK10,A2}. 
Using dopant-level Trojans~\cite{BRPB13, KJBP14}, it is even possible
to make a Trojan circuit look physically the same as a Trojan-free circuit.

There are three standard defenses against hardware Trojans~\cite{BHBN14}:
(1) detect Trojans with logic testing or by comparing chip side-channel information (e.g. power and temperature) to that
of a reference chip~\cite{AB07, YM08, PN09, NWDRS11, BGV15}, 
(2) monitor runtime behavior using sensors to detect dramatic changes in the chip's behavior (e.g. signal propagation delay) \cite{LL08, FBS13}, and
(3) design the hardware to make it more difficult to insert Trojans and
easier to detect them~\cite{bisa, harpoon, IEGT13, RKM08, SHWK11, fanci}.
Cryptographic techniques, such as verifiable computation 
protocols, can audit a chip's behavior~\cite{zebra,verifiable-comp,GKR08,trueset,pinocchio,adsnark,BCTV14,BCCT12,GGPR13}.
 
A backdoored U2F token could exfiltrate a user's cryptographic secrets
by encoding them in innocent-looking protocol messages. 
Simmons introduced the notion of subliminal channels~\cite{Simm84} to capture
this class of attacks and Desmedt proposed ``subliminal-free signatures''~\cite{D88} as a defense. 
Later work extended the notion of subliminal-freeness to zero-knowledge protocols
and other primitives~\cite{Simm84,Simm85,Des90,Des94,BDI+96,BBS98,OO90}.

Work on kleptography~\cite{YY97,RTYZ16,RTYZ17} and
algorithm-substitution attacks~\cite{BPR14,FM18}
models and defends against more general malicious implementations of 
cryptographic primitives.

Cryptographic reverse firewalls~\cite{MSD15} are a general technique
for defending against backdoored implementations of interactive cryptographic protocols.
A reverse firewall sits between a potentially backdoored
implementation and the outside world, modifying messages in such a way that (a)
preserves the security of the original protocol and (b) prevents the malicious
implementation from leaking information to the outside world.
In this work, we essentially build a reverse firewall for the special case
of U2F tokens.

Ateniese et al.~show that if 
a digital signature scheme $\Sigma$ is unique or rerandomizable
then there are very efficient reverse firewalls for $\Sigma$~\cite{AMV15}.
Our firewalled signing protocol is a reverse firewall for ECDSA, 
which is neither unique nor rerandomizable.
Our signing protocol, however, is interactive,
while theirs is not.

Recent work explores two-party protocols for ECDSA signing 
when each party holds a share of the signing key~\cite{L17, DKLS18}.
In contrast, in our setting the token has the entire secret key and
the browser's role is to enforce that the token samples its signing randomness
from the correct distribution.
The problem of enforcing good use of randomness also appears in 
prior work on collaborative key-generation
protocols~\cite{randkeys,JG02,controlled-randomness}.

\name uses a log-structured counter design.
Log-structured file systems go back to the work of
Rosenblum and Ousterhout~\cite{RO92}, and similar ideas
appear in flash-oriented filesystems~\cite{jffs, yaffs, elf, tffs},
key-value stores~\cite{fawn, flashstore, silt, skimpystash},
and other data structures~\cite{flashdb, fdtrees, bufferhash, microhash, bftl,
capsule}, many of which are tailored to
embedded devices. 
While much of this work focuses on building general file systems on flash, we
seek extreme space efficiency by tailoring our design to the very specific
needs of increment-only counters.

With the rise of browser protections against cookie
abuses, advertisers and trackers turned to novel fingerprinting techniques, such as 
the installation of browser plugins~\cite{NKJKPV13,NKJKPV15, 1mtracking}. 
U2F token fingerprinting gives advertisers yet another way to track
users across origins~\cite{w3c-fingerprint, RFC6973}.
The defenses we introduce in \name 
reduce the fingerprinting surface introduced by U2F.

\section{Conclusions}
\name shows that it \textit{is} possible to implement very strong defenses
against hardware Trojans for one important class of hardware devices.
Furthermore, the \name protections maintain server-side backwards
compatibility and come with unnoticeable performance overhead. 

In future work, we hope to extend the \name token design
to handle post-quantum signature schemes---such as those based on 
lattices or hash functions~\cite{PQreport,newhope,frodo,sphincs,ducas2018crystals}.
A second task is to add a notion of third-party auditability to \name.
That is, if the browser ever outputs ``Token failure,'' it should also output
a third-party verifiable proof of the token's misbehavior.
An honest token should be able to generate a similar proof if
the browser ever misbehaves.
Is it possible to achieve this stronger notion of security without
sacrificing performance?

\paragraph{Acknowledgements.}
We thank Ben Riva for pointing 
out an error in the proof of
Theorem~\ref{thm:vif-main} and Dmitry Kogan
for reviewing the corrected proof.
We would like to thank Bryan Ford and
Philipp Jovanovic for their thoughts on 
applications and extensions of \name.
Phil Levis pointed us to relevant related 
work on data structures optimized for flash-storage. 
Saba Eskandarian and Dmitry Kogan gave 
comments that improved the writing.
Marius Schilder provided implementation guidance
and helped us understand flash hardware constraints.

This work received support from NSF, DARPA, ONR, the Simons
Foundation, and CISPA.  Opinions, findings and conclusions or
recommendations expressed in this material are those of the authors
and do not necessarily reflect the views of DARPA.

{\small
\setstretch{0.95}
\bibliography{refs}
\bibliographystyle{abbrv} 
}

\appendix
\section{Selective failure attacks}
\label{app:fail}

In our security definitions (Section~\ref{sec:goals:sec}), 
we restrict our attention to tokens that never cause an honest browser to
abort.
We explain here why this restriction is not too severe.

Informally, say that an adversarial relying party $\Arp$ wants to collude with a
malicious token $\Atoken$ to achieve some goal $\calG$.
For example, a malicious token might want to leak its cryptographic
signing key to the malicious relying party.
We claim that if the adversary $(\Arp, \Atoken)$ achieves goal $\calG$ with
probability at most~$\epsilon$ when interacting with an adversarial
token that \textit{never} causes the honest browser to abort, then
the adversarial relying party achieves goal $\calG$ with probability
at most $\eabort \leq (T+1)\epsilon$, after at most $T$ registration or authentication
interactions with a token that possibly causes an abort at a chosen time.

Thus, if we can show that the probability of achieving $\calG$
is negligible when the token never causes the honest browser to abort, 
then this probability remains negligible even if the token triggers
an abort at an arbitrary time.
So a malicious token cannot use selective failure
to exfiltrate a cryptographic key, unless such an attack were possible
without selective failure.

To prove the claim informally: 
Assume that there exists an adversarial relying party and token
$\calA = (\Arp, \Atoken)$ such that 
(a)~$\Atoken$ possibly triggers a browser abort and 
(b)~$\calA$ achieves goal $\calG$ with probability $\eabort$.
Then we we can construct an adversary $\calA' = (\Atoken', \Arp')$ such that
(a)~$\Atoken'$ never triggers a browser abort and 
(b)~$\calA'$ achieves goal $\calG$ with probability $\epsilon = \eabort/(T+1)$.

The token $\Atoken'$ runs $\Atoken$, except that on the request
for which $\Atoken$ would have triggered 
a browser abort, $\Atoken'$ executes the protocol faithfully.
The relying party $\Arp'$ just guesses the index of the interaction on which $\Atoken$ would have
caused an abort (or that $\Atoken$ never causes an abort).
This guess is correct with independent probability $1/(T+1)$, so the advantage
of $\calA'$ is at least $\epsilon = \eabort/(T+1)$.

This argument crucially relies on the fact that once a token fails, the user
discards the token (i.e., the token can only abort \textit{once}), 
so it is important that the browser interface prevent
a user from continuing to use a failing token.

\section{The ECDSA signature scheme}
\label{app:ecdsa}

\newcommand{\Hmsg}{H_\textsf{msg}}
\newcommand{\Hg}{H_{\G}}
We follow the concise description of ECDSA of Fersch, Kiltz, and
Poettering~\cite{FKP16}.
The ECDSA signature scheme over message space $\calM$
uses a fixed group $\G = \langle g \rangle$ of prime order $q$.
For our applications, $\G$ is the NIST P256 elliptic curve group.
The scheme also uses a hash function $H: \calM \to \Z_q$ and 
a ``conversion function'' $f: \G \to \Z_q$.

The algorithms of the signature scheme are:
\begin{itemize}
  \item $\EKeyGen() \to (\sk, \pk)$. 
        Sample $x \getsr \Z_q$.
        Output $x$ as the secret key and $X=g^x \in \G$
        as the public key.
  \item $\ESign(\sk = x, m) \to \sigma$.
        \begin{itemize}
          \item Choose $r \getsr \Z^*_q$.
          \item Compute $c \gets f(g^r) \in \Z_q$.
          \item Compute $s \gets (H(m) + c \cdot x)/r \in \Z_q$.
          \item If $c = 0$ or $s = 0$,
                output $\bot$
                (i.e., fail). 
          \item Output $\sigma = (c, s)$.
        \end{itemize}
        We write $\ESign(\sk, m; r)$ to denote the deterministic operation
        of running $\ESign$ using randomness $r \in \Z^*_q$.

  \item $\EVerify(\pk, m, \sigma) \to \zo$.
        \begin{itemize}
          \item Parse the verification key $\pk$ as a group element $X \in \G$.
                Parse the signature $\sigma$ as a pair $(c,s) \in \Z_q^2$.
          \item If $c = 0$ or $s = 0$, output ``$0$.''
          \item Compute the value 
\begingroup
\setlength{\belowdisplayskip}{0pt} \setlength{\belowdisplayshortskip}{0pt}
\setlength{\abovedisplayskip}{0pt} \setlength{\abovedisplayshortskip}{0pt}
                \begin{align}
                  \wR \gets (g^{H(m)} X^c)^{1/s} \quad \in \G.\label{eq:ecdsa}
                \end{align}
\endgroup
          \item Output ``$1$'' iff 
            $\wR \neq 1 \in \G$ and $f(\wR) = c$.
        \end{itemize}
\end{itemize}

In practice, $H$ is a cryptographic hash function (e.g., SHA) and $f$
is the function that interprets a group element $g \in \G$ as an elliptic-curve
point and outputs the value of the $x$-coordinate of this point modulo $q$.

\paragraph{ECDSA signatures are malleable.}
ECDSA signatures are \textit{malleable}: that is, given a valid signature
$\sigma$ on $m$ under public key $X$, anyone can produce another valid
signature $\bar \sigma \neq \sigma$ on the same message $m$ under the same public key $X$.
The reason is that if $\sigma = (c,s) \in \Z_q^2$ is a valid signature, then
$\bar \sigma = (c, -s) \in \Z_q^2$ is also a valid signature.

\paragraph{Idealized ECDSA.}
We prove that the VIF construction of Section~\ref{sec:crypto:vif}
has $\Sigma$-pseudorandomness
and $\Sigma$-unforgeability when $\Sigma$ is an ``Idealized'' variant
of the ECDSA signature scheme.
Following Brickell et al.~\cite{Brickell}, we idealize ECDSA in the sense that
we model the two hash functions used in the ECDSA signature scheme as random
oracles~\cite{BR93}.
Modeling the conversion function $f$ as a random oracle is somewhat
problematic, as $f$ satisfies a number of properties that a truly
random function does not.
For example, for any $X \in \G$, $f(X) = f(1/X)$ 
(using multiplicative notation for the group operation).
Unfortunately, the peculiarities of ECDSA 
require resorting to some sort of idealization~\cite{Brown05,FKP16}.

\section{VIF security analysis}
\label{app:sig}

The experiments and algorithms in this section are are all implicitly
parameterized by a security parameter $\lambda$ and we require that all algorithms
run in probabilistic polynomial time in the parameter~$\lambda$.

\subsection{Standard definitions and preliminaries}
\paragraph{Discrete-log problem.}
Let $\G$ be a finite cyclic group of prime order $q$ generated by $g \in \G$.
We define the discrete-log advantage of an adversary $\calA$ as:
\[ \DLAdv[\calA, \G] \deq \Pr_{x \getsr \Z_q}\big[x = \calA(g, g^x)\big]. \]
We say that the discrete-log problem is hard in the group $\G$ if for all 
efficient adversaries $\calA$ (i.e., running in time polylogarithmic in the
group order $q$), $\DLAdv[\calA, \G]$ is negligible in $\log q$.

\paragraph{Verifiable random functions.}
We first recall the standard definition of pseudorandomness for a 
verifiable random function~\cite{VRF,DY05,NZ15}.
To do so, we define an experiment between a challenger
and an adversary. 
For $b \in \zo$,
let $\Wb{exp:vrf}$ denote the probability that
Experiment~\ref{exp:vrf} of Figure~\ref{fig:VRF}, parameterized with bit $b$, outputs the value ``$1$.''
Then define the advantage of an adversary $\calA$ 
at attacking the pseudorandomness of a VRF scheme $\calV$ as:
$\VRFAdv[\calA, \calV] \deq \abs{ \Wz{exp:vrf} - \Wo{exp:vrf} }$.
We say that a VRF scheme $\calV$ is secure
if, for all efficient adversaries $\calA$, 
$\VRFAdv[\calA, \calV]$ is negligible in the security parameter.

We also define a natural notion of unpredictability for VRFs.
Informally, we say that a VRF is resistant to ``set-targeting'' attacks if
the adversary cannot find a VRF input $x^* \in \calX$ that causes
the output of the VRF on that point to fall into a subset $\calY^* \subseteq \calY$,
except with probability $\abs{\calY^*}/\abs{\calY}$, plus a negligible term.

More formally, let $\Win{exp:vrfset}$ denote the probability that
Experiment~\ref{exp:vrfset} of Figure~\ref{fig:VRFSet}, outputs the value ``$1$.''
Then define the set-targeting advantage of an adversary $\calA$ 
at attaching a VRF scheme $\calV$ mapping
inputs in finite set $\calX$ to outputs in finite set $\calY$ as:
\[ \VRFSetAdv[\calA, \calV] \deq \abs{ \Win{exp:vrfset} - \frac{\abs{\calY^*}}{\abs{\calY}} },\]
where $\abs{\calY^*}$ is the size of the set (the ``target set'') that the adversary sends to 
the challenger in Experiment~\ref{exp:vrfset}.

\begin{figure}
  {\small
\begin{framed}
\noindent
\Experiment{VRF Pseudorandomness}\label{exp:vrf}
The experiment is parameterized by a VRF $\calV = (\VRFKeyGen,\allowbreak \VRFEval,\allowbreak \VRFVerify)$ over finite input space $\calX$ and finite output space $\calY$,
a bit $b \in \zo$, 
and an adversary $\calA$.
The experiment is an interaction between a challenger and the adversary $\calA$, 
and it proceeds as follows:
\begin{enumerate}
\item The challenger generates a VRF keypair $(\sk, \pk) \gets \VRFKeyGen()$.
      The challenger sends the public key $\pk$ to the adversary.
\item The challenger gives the adversary oracle access
      to the function $\VRFEval(\sk, \cdot)$.
\item At some point, the adversary $\calA$ sends a challenge 
      value $x^* \in \calX$ to the challenger.
      The adversary must not have queried the challenger at the point $x^*$.
  \begin{itemize}
    \item If $b = 0$, the challenger runs $(y^*, \pi) \gets \VRFEval(\sk, x^*)$,
      and sends $y^*$ to the adversary.
    \item If $b = 1$, the challenger samples $y^* \getsr \calY$.
      and sends $y^*$ to the adversary.
  \end{itemize}
\item The adversary may continue to make queries to $\VRFEval(\sk, \cdot)$, 
      provided that the adversary never queries the point $x^*$.
\item Finally, the adversary outputs a guess $\hat b$ of the bit $b$,
      and we define $\hat b$ to be the output value of the experiment.
\end{enumerate}
\end{framed}
}
\caption{VRF pseudorandomness experiment.} \label{fig:VRF}
\end{figure}

\begin{figure}
  {\small
\begin{framed}
\noindent
\Experiment{VRF set targeting}\label{exp:vrfset}
The experiment is parameterized by a VRF $\calV = (\VRFKeyGen,\allowbreak \VRFEval,\allowbreak \VRFVerify)$ 
over finite input space $\calX$ and finite output space $\calY$,
and an adversary $\calA$.
The experiment is an interaction between a challenger and the adversary $\calA$, 
and it proceeds as follows:
\begin{enumerate}
\item The challenger generates a VRF keypair $(\sk, \pk) \gets \VRFKeyGen()$.
      The challenger sends the public key $\pk$ to the adversary.
\item The challenger gives the adversary oracle access
      to the function $\VRFEval(\sk, \cdot)$.
\item At some point, the adversary chooses a ``target set'' 
      $\calY^* \subseteq \calY$ of size polynomial in the (implicit) security parameter
      and sends $\calY^*$ to the challenger.
      The adversary then may continue making queries to $\VRFEval(\sk,\cdot)$. 

\item Finally, the adversary outputs a value $x^* \in \calX$.
\item We say that the output of the experiment is ``$1$'' if 
      (1) the adversary never queried the VRF challenger on the point $x^*$ and 
      (2) $\VRFEval(\sk, x^*) \in \calY^*$. 
      The output of the experiment is ``$0$'' otherwise.
\end{enumerate}
\end{framed}
}
\caption{VRF set-targeting experiment.} \label{fig:VRFSet}
\end{figure}

\begin{lemma}\label{lemma:vrfset}
Let $\calV$ be a VRF with output space $\calY$.
For every efficient algorithm $\calA$ attacking $\calV$
in the set-targeting experiment using 
$Q$ queries and a target set of size $\abs{\calY^*}$,
there exists an efficient algorithm $\calB$ that 
attacks the VRF $\calV$'s pseudorandomness property using $Q$ queries such that:
\[ \VRFSetAdv[\calA, \calV] \leq \VRFAdv[\calB, \calV] + \frac{\abs{\calY^*}}{\abs{\calY}}.\]
Furthermore, $\calB$ runs in time linear in the running time of~$\calA$.
\end{lemma}

\begin{proof}
We first construct the algorithm $\calB$ and then prove that it satisfies the
claimed advantage.
Algorithm $\calB$ operates as follows:
  \begin{itemize}
    \item Receive the VRF public key $\pk$ from the VRF pseudorandomness challenger and
          forward $\pk$ to the adversary $\calA$.
    \item Run $\calA$ and forward all of $\calA$'s VRF queries to the VRF 
          pseudorandomness challenger.
    \item When the adversary $\calA$ outputs a target set $\calY^*$, store this set.
    \item When the adversary $\calA$ outputs its point $x^* \in \calX$:
          \begin{itemize}
              \item Forward $x^*$ to the VRF challenger as the challenge query.
              \item Receive a value $y^* \in \calY$ from the VRF challenger.
              \item If $y^* \in \calY^*$, output ``$1$.'' 
                    Otherwise, output ``$0$.''
          \end{itemize}
  \end{itemize}

To argue that $\calB$ achieves the claimed advantage:
When interacting with the VRF (the VRF challenger's bit $b=0$), the probability
that the experiment outputs ``$1$'' is exactly the probability that the adversary
wins in the VRF set-targeting experiment.
When interacting with the random function (the VRF challenger's bit $b=1$),
the probability that the experiment outputs ``$1$'' is the probability that
  a uniform random value from $\calY$ falls into $\calY^*$, which is $\abs{\calY^*}/\abs{\calY}$.
  Therefore, 
  \[ \VRFAdv[\calB, \calV] \geq \abs{\VRFSetAdv[\calA, \calV] - \frac{\abs{\calY^*}}{\abs{\calY}}}, \]
which proves the lemma.
\end{proof}

\subsection{VIF security definitions}
\label{app:sig:vifdefs}

In this section, we give the complete definition of
security for the \textit{verifiable identity family}
primitive introduced in Section~\ref{sec:crypto:vif}. 

\begin{figure}[t]
  {\small
\begin{framed}
\noindent
\Experiment{VIF $\Sigma$-pseudorandomness}\label{exp:vifpr}
The experiment is parameterized by
a signature scheme $\Sigma = (\SigKeyGen,\allowbreak \SigSign,\allowbreak \SigVerify)$
  with message space $\calM$,
a VIF scheme $\Phi_\Sigma = (\VIFKeyGen, \VIFEval, \VIFVerify)$
with identity space $\ID$, a bit $b \in \zo$, and 
an adversary~$\calA$.

The experiment is an interaction between a challenger and the adversary $\calA$, 
and it proceeds as follows:
\begin{enumerate}
\item The challenger runs $(\msk, \mpk) \gets \VIFKeyGen()$.
      and sends $\mpk$ to the adversary.
    \item The challenger gives the adversary access to an \textbf{identity oracle}
      $\Oid$ and a \textbf{signature oracle} $\Osig$, defined as follows:
      \begin{itemize}
        \item $\Oid(\id \in \ID):$
          \begin{itemize}
            \item $(\sk_\id, \pk_\id, \pi) \gets \VIFEval(\msk, \id)$
            \item Return $(\pk_\id, \pi)$
          \end{itemize}

        \item $\Osig(\id \in \ID, m \in \calM):$
          \begin{itemize}
            \item $(\sk_\id, \pk_\id, \pi) \gets \VIFEval(\msk, \id)$
            \item Return $\SigSign(\sk_\id, m)$
          \end{itemize}
          
      \end{itemize}

\item At some point, the adversary $\calA$ sends an identity $\id^*$ to the challenger.
      The adversary must not have previously queried the identity or signature oracles
      at $\id^*$. 

  \begin{itemize}
    \item If $b = 0$, the challenger runs $(\sk_{\id^*}, \pk_{\id^*}, \pi) \gets \VIFEval(\msk, \id^*)$,
      and sends $\pk_{\id^*}$ to the adversary.
    \item If $b = 1$, the challenger runs $(\sk_\ms{rand}, \pk_\ms{rand}) \gets \SigKeyGen()$,
        and sends $\pk_\ms{rand}$ to the adversary.
  \end{itemize}
\item The adversary may continue to make identity- and signature-oracle queries,
      provided that the adversary never makes an identity query using identity $\id^*$.
      The adversary may make subsequent signing queries for identity $\id^*$:
      \begin{itemize}
        \item If $b=0$, the challenger signs using $\sk_{\id^*}$.
        \item If $b=1$, the challenger signs using $\sk_\ms{rand}$.
      \end{itemize}
\item Finally, the adversary outputs a guess $\hat b$ for $b$,
  and the value $\hat b$ is the output of the experiment.
\end{enumerate}
\end{framed}
  }
\caption{VIF pseudorandomness experiment.} \label{fig:PR}
\end{figure}

\paragraph{Defining pseudorandomness.}
We define pseudorandomness using Experiment~\ref{exp:vifpr}, shown in Figure~\ref{fig:PR}. 
For a bit $b \in \zo$, 
let $\Wb{exp:vifpr}$ denote the probability that
the output of Experiment~\ref{exp:vifpr}, with challenge bit $b$, outputs ``$1$.''
Then define the advantage of an adversary $\calA$ 
at attacking the pseudorandomness of a VIF scheme $\Phi_\Sigma$ as:
$\PRAdv[\calA, \Phi_\Sigma] \deq \abs{ \Wz{exp:vifpr} - \Wo{exp:vifpr} }$.
A VIF scheme $\Phi_\Sigma$ satisfies \textit{$\Sigma$-pseudorandomness} 
if for all efficient adversaries $\calA$, 
$\PRAdv[\calA, \Phi_\Sigma]$ is negligible in the security parameter.

\begin{figure}
{\small 
\begin{framed}
\noindent
\Experiment{VIF $\Sigma$-unforgeability}\label{exp:vifuf}
The experiment is parameterized by
a signature scheme $\Sigma$ with message space $\calM$,  
and a VIF scheme $\Phi_\Sigma = (\VIFKeyGen, \VIFEval, \VIFVerify)$
with identity space $\ID$, and 
an adversary~$\calA$.

The experiment is an interaction between a challenger and the adversary $\calA$, 
and it proceeds as follows:
\begin{enumerate}
\item The challenger runs $(\msk, \mpk) \gets \VIFKeyGen()$.
      and sends $\mpk$ to the adversary.
\item The challenger gives the adversary access to 
      an \textbf{identity oracle} and a \textbf{signature oracle},
      as in Experiment~\ref{exp:vifpr}.
\item The adversary $\calA$ outputs a tuple $(\id^*, m^*, \sigma^*)$,
      where $\id^* \in \ID$ $m^* \in \calM$.
\item The output of the experiment is the bit $b = 1$ if
      \begin{itemize}
        \item the adversary never queried the signature oracle on the pair
              $(\id^*, m^*)$, and
        \item $\sigma^*$ is a valid signature on $m^*$ using the public key
          corresponding to identity $\id^*$.
              That is, if $(\sk_{\id^*}, \pk_{\id^*}, \pi) \gets
                \VIFEval(\msk, \id^*)$ and ${\SigVerify(\pk_{\id^*}, m^*, \sigma^*) = 1}$.
      \end{itemize}
      The output is $b=0$ otherwise.
\end{enumerate}
\end{framed}
}
\caption{VIF unforgeability experiment.} \label{fig:UF}
\end{figure}

\paragraph{Defining unforgeability.}
We define unforgeability using Experiment~\ref{exp:vifuf} shown in Figure~\ref{fig:UF}. 
Define the $\Sigma$-Unforgeability advantage of an adversary $\calA$ 
at attacking a VIF scheme $\Phi_\Sigma$
as the probability that Experiment outputs ``$1$.'' 
Denote this probability as $\UFAdv[\calA, \Phi_\Sigma]$.
A VIF scheme $\Phi_\Sigma$, defined relative to a signature
scheme $\Sigma$, satisfies \textit{$\Sigma$-unforgeability} 
if for all efficient adversaries
$\calA$, $\UFAdv[\calA, \Phi_\Sigma]$ is negligible in the security parameter.

\subsection{VIF security proofs}
\label{app:sig:proofs}

\begin{theorem}\label{thm:vifpr}
The VIF construction of Section~\ref{sec:crypto:vif}, 
satisfies $\Sigma$-pseudorandomness when $\Sigma$
is the Idealized ECDSA signature scheme of Appendix~\ref{app:ecdsa}
and the VIF is instantiated with a secure VRF. 

In particular, let $\Fdsa$ be the VIF construction of
Section~\ref{sec:crypto:vif} instantiated with 
cyclic group $\G$ of prime order $q$ and a VRF $\calV$
with output space $\Z^*_q$.
Then, for every efficient algorithm $\calA$ attacking 
the VIF pseudorandomness of $\Fdsa$
using at most 
$\Qsig$ signing-oracle queries and $\Qid$ identity-oracle queries 
there exists an efficient algorithm $\calB$ attacking $\calV$
using at most $\Qid + \Qsig$ VRF queries, 
such that: 
\[\PRAdv[\calA, \Fdsa] \leq \VRFAdv[\calB, \calV].\]
\end{theorem}

\begin{proof}
From an adversary $\calA$ that breaks the $\Sigma$-pseudorandomness
of the VIF, we construct an adversary $\calB$
that breaks the pseudorandomness of the VRF.
Let $g \in \G$ be a generator of the group $\G$ of prime order $q$
over which ECDSA is defined.
The algorithm $\calB$ works as follows:
\begin{itemize}
  \item Receive a VRF public key $\pkVRF$ from the VRF challenger.
  \item Generate an ECDSA signing key $x \getsr \Z_q$
        and the corresponding public key $X = g^x \in \G$.
        Then, send $(X, \pkVRF)$ to $\calA$ as the VIF master public key.
  \item Algorithm $\calA$ makes identity and signing queries.
        \begin{itemize}
          \item For each identity query $\id$ that $\calA$ makes, 
                query the VRF challenger on $\id$
                and receive a response $(y, \piVRF)$.
                Respond to the query with the pair $(\pk_\id, \pi)$,
                where $\pk_\id \gets X^y$ and $\pi \gets (y, \piVRF)$.
          \item For each signing query $(\id, m)$ that $\calA$ makes, 
                query the VRF challenger on identity $\id$, 
                and receive a pair $(y, \pi)$.
                Then, compute $\sk_\id \gets x \cdot y \in \Z_q$
                and $\sigma \gets \SigSign(\sk_\id, m)$,
                and return $\sigma$ to~$\calA$.
        \end{itemize}
  \item Upon receiving the VIF challenge point $\id^*$ from $\calA$, 
        forward this value to the VRF challenger.
        Receive the VRF challenge $y^*$ from the VRF challenger,
        set $\sk_{\id^*} \gets x \cdot y^* \in \Z_q$,
        and send $Y^* \gets g^{x \cdot y^*} \in \G$ to the adversary $\calA$.

        If $\calA$ makes a signing query for message $m$ on identity $\id^*$, 
        respond with $\sigma \gets \SigSign(\sk_{\id^*}, m)$.

  \item Output whatever $\calA$ outputs.
\end{itemize}

When running $\calB$ in the VRF experiment with
VRF $\calV$ and challenge bit $b$, $\calA$'s
view is exactly as in Experiment~\ref{exp:vifpr}
with VRF $\calV$ and challenge bit $b$.
Thus, $\calB$ achieves the same advantage at attacking
the VRF $\calV$ as $\calA$ does in attacking 
the VIF's ECDSA-pseudorandomness property.
\end{proof}

\newcounter{GameCounter}
\setcounter{GameCounter}{-1}
\newcommand{\game}{\refstepcounter{GameCounter}\paragraph{Game~\arabic{GameCounter}.}}
\newcommand{\gref}[1]{Game~\ref{#1}}
\newcommand{\Wgame}[1]{W_{\ref{#1}}}

\paragraph{Alternative VIF constructions that are insecure.}
Before proving the unforgeability of our VIF construction, we give
some intuition for why we construct the VIF the way we do.

First, if we remove the verifiable random function (VRF) from our
VIF construction, it is insecure.
That is, we might consider letting $\Z_q$ be the space of identities
and letting the signature under identity $\id \in \Z_q$ with master
secret key $x \in \Z_q$ be an ECDSA signature under secret 
key $x \cdot \id \in \Z_q$ or $x + \id \in \Z_q$.

As Morita et al.~observe~\cite{RKA15} in the context of related-key
attacks on ECDSA, both of these constructions are insecure.
\begin{itemize}
  \item \textit{Multiplicative variant.}
        In this insecure VIF construction, the signature with master secret key 
        $x \in \Z_q$ under identity $\id \in \Z_q$ is an ECDSA signature under
        secret key $\sk_\id \gets x \cdot \id \in \Z_q$.

        To produce a forgery, the adversary:
        \begin{itemize}
          \item requests a signature on some message $m_0$ under some identity~$\id_0$,
          \item chooses an arbitrary message $m_1$ such that~$m_1 \neq m_0$,
          \item computes 
                \begin{align*}
                  \id_1 &\gets \id_0 \cdot H(m_1) / H(m_0) &\in \Z^*_q\\
                    s_1 &\gets s_0 \cdot H(m_1) / H(m_0)  &\in \Z^*_q
                \end{align*}
          \item and outputs $(\id_1, m_1, (c_0, s_1))$ as its forgery.
        \end{itemize}

  \item \textit{Additive variant.}
        In this insecure VIF construction, the signature with master secret key 
        $x \in \Z_q$ under identity $\id \in \Z_q$ is an ECDSA signature under
        secret key $\sk_\id \gets x + \id \in \Z_q$.
        To produce a forgery, the adversary:
        \begin{itemize}
          \item requests a signature on some message $m_0$ under some identity~$\id_0$,
          \item chooses an arbitrary message $m_1$ such that~$m_1 \neq m_0$,
          \item computes 
                \[ \id_1 \gets \id_0 + (H(m_0) - H(m_1))/c_0 \quad \in \Z^*_q.\]
          \item and outputs $(\id_1, m_1, (c_0, s_0))$ as its forgery.
        \end{itemize}
\end{itemize}

There are at least two ways to protect against these attacks.
The first is to constrain the identity space to be a subset
$\calI \subseteq \Z_q$ such that $|\calI| \ll q$.
For example, we could let $\calI$ be the set of values $\id \in \Z_q$ 
such that $|\id| \leq \sqrt{q}$.
Constraining the identity set in this way
defeats the attacks above because the forged identities produced
in the attacks above will be distributed essentially uniformly over $\Z_q$
and if $|\calI| \ll q$, the forged identity will rarely fall into the set $\calI$
of valid identities.

The second way to defeat the attacks is to essentially
randomize the adversary's choice of identity.
In our construction, this is achieved through the use of verifiable random
functions, which we anyhow need to provide the VIF pseudorandomness 
property (Theorem~\ref{thm:vifpr}).
When using our VRF-based construction, the attacker may request signatures
under secret keys $x_\id = x + \VRFEval(\skVRF, \id)$, for identities $\id$ of the
attacker's choosing.
However, when instantiating this construction with a secure VRF, the 
probability that the VRF ever outputs a ``bad'' value that enables
the adversary to produce one of the forgeries as above is negligible.
We give a detailed argument in the security analysis below.

\paragraph{Preliminaries to VIF unforgeability proof.}
The following notion will be convenient for our proof of VIF unforgeability.
\begin{defn}[Useful query]\label{defn:useful}
Let $\calA$ be an attacker in the VIF $\Sigma$-unforgeability game,
where $\Sigma$ is the Idealized ECDSA signature scheme of Appendix~\ref{app:ecdsa}.
Informally, we say that $\calA$ makes a \emph{useful query} in its interaction
with the challenger if the signing-oracle and $f$-oracle queries that
the adversary receives reveal enough information about the oracle $f$ for
the adversary to conclude that its forged signature is valid.

Formally, $\calA$ makes a \emph{useful query} if 
there exists a value $R \in \G$ such that
\begin{itemize}
\item either:
      \begin{itemize}
        \item the adversary queries its $f$-oracle at the point $R$, or
        \item the adversary makes a signing-oracle query whose response
              implicitly defines the value $f(R)$, and
      \end{itemize}
\item the adversary outputs a forgery $(\id, m, (c,s))$ such that
      $R = (g^{H(m)} X^{yc})^{1/s} \in \G$, where 
      \begin{itemize}
        \item $H(m)$ is the value of the challenger's $H$-oracle evaluated
              at the point $m$, 
        \item $X$ is the challenger's ECDSA master public key, and
        \item $y \gets \VRFEval(\skVRF, m)$, where $\skVRF$ is the challenger's
              VRF secret key.
    \end{itemize}
\end{itemize}
\end{defn}

\begin{defn}[Adversary in standard form]\label{defn:standard}
We say that a VIF unforgeability adversary $\calA$ is in 
\emph{standard form} if it:
\begin{itemize}
  \item makes an $H$-query on message $m$ before making a signing-oracle 
        query on $m$ or outputting a forgery on $m$, 
  \item makes an identity query on $\id$ before making a signing-oracle
        query under identity $\id$ or outputting a forgery under identity $\id$, and 
  \item always makes a useful query, in the 
        sense of Definition~\ref{defn:useful}.
\end{itemize}
\end{defn}
It is possible to convert any adversary $\calA$ into one in standard form by
only slightly increasing its query complexity.
That is, given an arbitrary VIF unforgeability adversary that makes:
\begin{itemize}
  \item $\Qid$ identity queries,
  \item $\Qsig$ signing queries, 
  \item $Q_f$ queries to its $f$-oracle, and
  \item $Q_H$ queries to its $H$-oracle
\end{itemize}
we can convert it to one in standard form that makes
\begin{itemize}
  \item $\Qid + \Qsig + 1$ identity queries,
  \item $\Qsig$ signing queries, 
  \item $Q_f + 1$ queries to its $f$-oracle, and
  \item $Q_H + \Qsig + 1$ queries to its $H$-oracle
\end{itemize}
Therefore, if a VIF construction is unforgeable against all efficient 
adversaries in standard form, it is unforgeable against all efficient adversaries.

\begin{theorem}\label{thm:vifuf}
The VIF construction of Section~\ref{sec:crypto:vif}
satisfies $\Sigma$-unforgeability when $\Sigma$
is the Idealized ECDSA signature scheme of Appendix~\ref{app:ecdsa},
when the VIF is instantiated with a secure VRF,
and when both $H$ and $f$ are modeled as random-oracles.

In particular, let $\Fdsa$ be the VIF construction of
Section~\ref{sec:crypto:vif} instantiated with the Idealized
ECDSA signature scheme over a group $\G$ of prime order~$q$
with a VRF~$\calV$.
Then, for every efficient algorithm $\calA$ in standard
form (Definition~\ref{defn:standard}) attacking $\Fdsa$
in the VIF unforgeability game using at most 
  \begin{itemize}
    \item $\Qid$ identity queries,
    \item $\Qsig$ signing queries, 
    \item $Q_f$ queries to its $f$-oracle, and
    \item $Q_H$ queries to its $H$-oracle
  \end{itemize}
there exists:
  \begin{itemize}
    \item  an efficient algorithm $\Bdlog$ attacking the discrete-log problem in~$\G$ and
    \item  an efficient algorithm $\Bvrf$ attacking the VRF $\calV$ using
           at most $2 \cdot \Qid$ queries
  \end{itemize}
such that: 
  \begin{multline*}
    \UFAdv[\calA, \Fdsa] \leq \\
    (Q_f + \Qsig) \cdot \sqrt{\epsilon_\ms{Dlog} 
        + \epsilon_\ms{VRF}}
        + \epsilon_{\ms{coll}_1} + \epsilon_{\ms{coll}_2},
  \end{multline*}
    where 
    \begin{align*}
      \epsilon_\ms{Dlog} &= \DLAdv[\Bdlog, \G]\\
      \epsilon_\ms{VRF}  &= \VRFAdv[\Bvrf, \calV] + (2 \cdot Q_H)/q\\
      \epsilon_{\ms{coll}_1} &= (Q_H + 1)/q\\
      \epsilon_{\ms{coll}_2} &= 2(Q_f + \Qsig)^2/q.
    \end{align*}
Furthermore, the algorithms $\Bdlog$ and $\Bvrf$ run 
in time linear in the running time of $\calA$.
\end{theorem}

\begin{proof}[Proof idea for Theorem~\ref{thm:vifuf}]
The proof of Theorem~\ref{thm:vifuf} works in three steps and
the outline loosely follows the security analysis of Schnorr
signatures presented in Boneh and Shoup~\cite{TheBook}.
\begin{enumerate}
\item First, we show in Lemma~\ref{lemma:uf-single} 
      that an adversary that (1) breaks the VIF unforgeability property,
      (2) makes a useful query
      (in the sense of Definition~\ref{defn:useful}), and
      (3) makes \textit{either} a single 
      $f$-oracle query \textit{or} a single signing-oracle 
      query implies an algorithm for breaking the 
      discrete-log problem in the group $\G$, 
      unless a certain failure event $F$ occurs.
\item Next, by appealing to the security of the VRF $\calV$,
      we show in Lemma~\ref{lemma:vrf} that the failure event $F$
      occurs with only negligible probability. 
      This proves that our VIF construction is secure against 
      adversaries that make a useful query and that make 
      only a single $f$-oracle or signing-oracle query.
\item Finally, we show in Lemma~\ref{lemma:many} how to convert an algorithm that breaks
      the VIF unforgeability property using an \emph{arbitrary} number of 
      $f$- or signing-oracle queries into one that makes a useful query
      and that makes at most a \emph{single} query to these two oracles in total. 
\end{enumerate}
Putting these three steps together completes the proof.
\end{proof}

Define $\UFAdvs[\calA, \Fdsa]$ to be the VIF unforgeability
advantage of algorithm $\calA$ at attacking $\Fdsa$ when the algorithm
$\calA$ is constrained to make at most \emph{either} a 
single useful signing-oracle query
\emph{or} a single useful $f$-oracle query.
(The algorithm $\calA$ may make any number of identity- or $H$-oracle queries.)

Then we first claim that:
\begin{lemma}\label{lemma:uf-single}
Let $\Fdsa$ be the VIF construction of
Section~\ref{sec:crypto:vif} instantiated with the Idealized
ECDSA signature scheme over a group $\G$ of prime order~$q$
with a VRF~$\calV$.
Let $\calA$ be a efficient algorithm $\calA$ attacking $\Fdsa$
that:
  \begin{itemize}
    \item makes at most $\Qid$ identity queries, 
    \item makes at most $Q_H$ $H$-oracle queries, 
    \item makes \emph{either} a single signing-oracle query 
          \emph{or} a single $f$-oracle query, and
    \item makes a useful query, 
          in the sense of Definition~\ref{defn:useful}.
  \end{itemize}
Then there exists:
  \begin{itemize}
    \item  an efficient algorithm $\Bdlog$ attacking the discrete-log problem in~$\G$ and
    \item  an efficient algorithm $\Bvrf$ attacking the VRF $\calV$ using
           at most $2\cdot \Qid$ queries
  \end{itemize}
such that: 
    \[ \UFAdvs[\calA, \Fdsa] \leq \sqrt{\epsilon_\ms{Dlog} 
        + \epsilon_\ms{VRF}}
        + \epsilon_{\ms{coll}_1},\]
    where $\epsilon_\ms{Dlog}$, $\epsilon_\ms{VRF}$, and
    $\epsilon_{\ms{coll}_1}$ are as in Theorem~\ref{thm:vifuf}.

Furthermore, the algorithms $\Bdlog$ and $\Bvrf$ run 
in time linear in the running time of $\calA$.
\end{lemma}

\begin{proof}
We prove the lemma by showing that if the adversary 
produces a valid VIF forgery after making a single
query to the $f$-oracle or signing oracle, we can
rewind the adversary and replay it to get a second
forgery relative to the \emph{same point} on the $f$-oracle.
Then, using these two valid forgeries, we can extract
the discrete log of the master ECDSA public key.

We prove the lemma with a sequence of games, defined
over a shared probability space,
in which the adversary interacts with a challenger.
We use $W_i$ to denote the event that the adversary
``wins'' in Game~$i$, where we will define
the winning event differently for each game.

\game\label{game:start}
In this game, the adversary $\calA$ interacts with the
VIF unforgeability challenger in the setting in which
the adversary may make at most a single $f$-oracle query 
or signing-oracle query.
The challenger responds to the adversary's $f$-oracle and 
$H$-oracle queries by lazily defining the values of these random oracles
and by remembering its past responses so that it can reply consistently. 

We implement the Game-\ref{game:start} challenger slightly differently
than the VIF challenger in that, if the adversary makes a signing-oracle
query, the challenger generates the signature in response 
by programming the $f$-oracle.
In other words, the challenger does not use the secret ECDSA signing
key to produce the response to the adversary's signing query.

That is, in response to a signing query on identity $\id$ and message $m$,
the challenger computes $y \gets \VRFEval(\skVRF, \id)$, chooses 
$c,s \getsr \Z^*_q$, and computes
\[ R \gets (g^{H(m)} \cdot X^{yc})^{1/s} \quad \in \G, \]
where $X$ is the challenger's master ECDSA public key.
(If $R$ is the identity element in $\G$, the challenger computes a new signature
with fresh randomness.)
The challenger then defines $f(R) \deq c$.
Since we only allow $\calA$ to make a single $f$-oracle or signing query,
the value $f(R)$ will always be undefined at the point at which the challenger
sets $f(R) \deq c$.

We must argue that the ECDSA signatures produced in this way are indistinguishable
to the adversary from signatures produced using the secret ECDSA signing key.
To do so, observe that a true signature $(c,s)$
is computed by sampling $r \getsr \Z^*_q$,
computing $c \gets f(g^r)$ and setting $s \gets (H(m) + xyc)/r \in \Z^*_q$,
where $x$ is the master ECDSA secret key.

The values $r$ and $c = f(g^r)$ are distributed independently and uniformly at random 
over $\Z^*_q$, over the choice of the random oracle $f$.
Each choice of the pair $(r,c)$ fixes a unique non-zero signature value $s$.
When we program the random oracle, we execute exactly the same 
sampling procedure just in a different
order: choose non-zero $c$ and $s$ at random from $\Z^*_q$ and
compute the unique $r$ (or $R = g^r$) that causes the signature to be valid
relative to the oracle $f$.
Therefore, the signatures that the Game-\ref{game:start} challenger
gives to the adversary are indistinguishable from ECDSA signatures
generated using the secret signing key.

We say that the adversary wins in \gref{game:start} if it outputs
a valid VIF forgery, according to the 
winning condition of Experiment~\ref{exp:vifuf}.

We have, by definition,
\begin{align}  
  \Pr[\Wgame{game:start}] = \UFAdvs[\calA, \Fdsa]. \label{eq:game0}
\end{align}

\game\label{game:zero}
In \gref{game:zero}, we modify the winning condition to require
that the output of the $H$-oracle is never zero.
We require this condition to hold for the proof of Lemma~\ref{lemma:vrf}.
Since the adversary can make at most $Q_H$ $H$-oracle queries, and since
each has a probability $1/q$ of being zero, we have
\[ \abs{\Pr[\Wgame{game:start}] - \Pr[\Wgame{game:zero}]} \leq Q_H/q. \]
(Since we have already assumed that the adversary is in standard form, as in
Definition~\ref{defn:standard}, answering the adversary's signing queries does 
not require fixing any more points in the $H$-oracle.)

\game\label{game:fork}
In \gref{game:fork}, we modify both the experiment and the winning
condition.
In particular, the challenger runs the adversary $\calA$ until it 
outputs a forgery.

\begin{itemize}
  \item If the adversary has made a signing-oracle query, the challenger
        does not run the adversary again.

  \item If the adversary has made an $f$-oracle query, 
        the challenger rewinds the adversary to the point after the adversary
        made its $f$-oracle query but before the challenger responded to 
        the adversary's query.

        Let $R \in \G$ be the adversary's $f$-oracle query.
        The challenger chooses a fresh response
        $c_1 \getsr \Z_q$ to the $f$-oracle query, redefines $f(R) \deq c_1$,
        and returns $c_1$ to the adversary as the
        response to its $f$-oracle query.
        The challenger then continues the execution of $\calA$ until 
        it outputs a forgery.
\end{itemize}

In the latter case, 
we can think of the challenger as ``forking'' the execution
of the algorithm $\calA$ at the point at which the challenger
responds to the adversary's $f$-oracle query.
We say that the first execution of $\calA$ after this forking
point is the ``left fork'' (in which $f(R) = c_0$)
and the second execution of $\calA$ after 
the forking point (in which $f(R) = c_1$)
is the ``right fork.''
In the two forks, whenever $c_0 \neq c_1$,
the adversary will be interacting with different $f$-oracles.

In \gref{game:fork}, the winning condition is as in 
\gref{game:start}, except that to win in \gref{game:fork},
the adversary must additionally either:
\begin{itemize}
  \item make a signing-oracle query, or
  \item make an $f$-oracle query, $c_0 \neq c_1$, and 
        the adversary must produce a forgery in both forks.
\end{itemize}

Let $\Pr[\Wgame{game:zero}] = \epsilon$ and
let $\Pr[\Wgame{game:fork}] = \epsilon'$.
By the Rewinding Lemma~\cite{TheBook},
$\epsilon' \geq \epsilon^2 - \epsilon/q$.
Assume, without loss of generality, that $\epsilon \geq 1/q$.
Then
\begin{align*}
  \epsilon' &\geq \epsilon^2 - \epsilon/q = \epsilon^2 - 2\epsilon/q + \epsilon/q\\
            &\geq \epsilon^2 - 2\epsilon/q + 1/q^2 = (\epsilon - 1/q)^2
\end{align*}
and we conclude that $\epsilon \geq \sqrt{\epsilon'} + 1/q$,
or that 
\[ \Pr[\Wgame{game:zero}] \leq \sqrt{\Pr[\Wgame{game:fork}]} + 1/q.\]

\game\label{game:dlog}
In \gref{game:dlog}, we modify the winning criteria of \gref{game:fork}
by adding an additional condition.

From the forgeries that the adversary produces in the two forks of the 
execution, we will be able to extract the discrete log of the ECDSA
master public key.
However, this extraction will fail if a certain linear relation holds
amongst the outputs of the adversary's identity- and random-oracle queries.
In \gref{game:dlog}, we argue that no efficient adversary can cause
this bad linear relation to hold often. 

\gref{game:dlog} proceeds as the prior game, except that at the conclusion
of the game, the challenger performs an additional check on the adversary's
forgeries.

We define two identity-message-signature triples:
\[ T_0 = (\id_0, m_0, (c_0, s_0))\quad\text{and}\quad T_1 = (\id_1, m_1, (c_1, s_1)).\]
The challenger defines these triples 
differently depending on the adversary's behavior:
\begin{itemize}
  \item If the adversary makes a signing-oracle query, then $T_0$ consists
        of the adversary's signing-oracle query $(\id_0, m_0)$ and the signing-oracle's
        response $(c_0, s_0)$ and $T_1$ consists of the adversary's left-fork forgery.
  \item If the adversary makes an $f$-oracle query, then 
        $T_0$ consists of the adversary's left-fork forgery and $T_1$ consists of 
        the adversary's right-fork forgery.
\end{itemize}

Using its VRF secret key $\skVRF$, the challenger computes:
\begin{align*}
  y_0 &\gets \VRFEval(\skVRF, \id_0)\\
  y_1 &\gets \VRFEval(\skVRF, \id_1)\\
  r_0 &\gets (H(m_0) + c_0 y_0 x ) / s_0 \quad \in \Z_q\\
  r_1 &\gets (H(m_1) + c_1 y_1 x ) / s_1 \quad \in \Z_q.
\end{align*}
(Observe that the inverses of $s_0$ and $s_1$ in $\Z_q$ are well
defined since these values are both components of valid ECDSA signatures.)

We say that the adversary wins in \gref{game:dlog} if it wins in \gref{game:fork}
and also 
\begin{align}
  \frac{c_0 y_0}{s_0} \neq \frac{c_1 y}{s_1} \quad \in \Z_q. \label{eq:goodrel}
\end{align}

In Lemma~\ref{lemma:vrf}, we
construct an efficient algorithm $\Bvrf$ attacking the VRF $\calV$ 
using at most $2\cdot \Qid$ VRF queries such that
 \[ \abs{ \Pr[\Wgame{game:fork}] - \Pr[\Wgame{game:dlog}]} \leq \VRFSetAdv[\Bvrf, \calV] + \frac{2\cdot Q_H}{q}.\] 

\paragraph{Final reduction.}
Now, we use the adversary $\calA$ of \gref{game:dlog} to 
construct an efficient algorithm $\Bdlog$ such that
\[ \Pr[\Wgame{game:dlog}] = \DLAdv[\Bdlog, \G].\] 

The adversary $\Bdlog$ executes the following steps:
\begin{itemize}
  \item Receive a pair $(g, X) \in \G^2$
        from the discrete-log challenger.
  \item Run $\calA$, playing the role of the \gref{game:dlog} challenger.
        To generate the VIF master public key, sample a fresh 
        VRF keypair 
        \[ (\skVRF, \pkVRF)\gets \VRFKeyGen()\]
        and set $\mpk \gets (X, \pkVRF)$,
        where $X \in \G$ is the value from 
        the discrete-log challenger.
        Send the VIF master public key $\mpk$ to~$\calA$.
  \item Answer $\calA$'s $H$-oracle and $f$-oracle queries using 
        a table.
  \item Answer $\calA$'s identity-oracle queries using the VRF secret
        key $\skVRF$.
  \item Answer $\calA$'s oracle signing query by programming the $f$-oracle
        as in Game~\ref{game:start}.
        In this way, the reduction can answer $\calA$'s signing query
        without knowing the secret ECDSA signing key (i.e., the discrete log
        of the challenge point $X \in \G$).
  \item Finally $\calA$ outputs forgeries $(\id_0, m_0, \sigma_0)$
        and $(\id_1, m_1, \sigma_1)$.
\end{itemize}

Now, whenever $\calA$ wins in Game~\ref{game:dlog}, algorithm $\Bdlog$
recovers two valid forgeries as in \gref{game:dlog},
\[ T_0 = (\id_0, m_0, (c_0, s_0)) \quad\text{ and }\quad T_1 = (\id_1, m_1, (c_1, s_1)),\]
and we define these two forgeries as in \gref{game:dlog}.
As in \gref{game:dlog}, define
\begin{align*}
  y_0 &\gets \VRFEval(\skVRF, \id_0)\\
  y_1 &\gets \VRFEval(\skVRF, \id_1).
\end{align*}

The algorithm $\Bdlog$ operates differently depending on whether
(I) makes a single signing-oracle query or
(II) the algorithm $\calA$ makes a single $f$-oracle query. 
We treat the two cases separately.

\itpara{Case I: Adversary makes a signing-oracle query.}
When the adversary makes a signing-oracle query, we can extract
a discrete log from (a) the response to the adversary's signing-oracle
query and (b) the adversary's left-fork forgery. 

Let $R \gets (g^{H(m_1)} X^{c_1 y_1})^{1/s_1} \in \G$.
Since the adversary makes a useful query, it must be that
$f(R)$ is defined at the end of the experiment.
Because the adversary may make at most one signing-oracle query,
the $f$-oracle is only defined at a single point $R$, and
the response to the adversary's signing-oracle query must satisfy:
\begin{align*}
  R = (g^{H(m_0)} X^{c_0 y_0})^{1/s_0} \in \G.
\end{align*}

Therefore, 
\begin{align*}
  \qquad(g^{H(m_0)} X^{c_0 y_0})^{1/s_0} &= (g^{H(m_1)} X^{c_1 y_1})^{1/s_1} &&\in \G&\\
  \intertext{which implies that}
  \qquad\frac{H(m_0)}{s_0} + \frac{c_0 y_0}{s_0}\cdot x &= \frac{H(m_1)}{s_1} + \frac{c_1 y_1}{s_1} \cdot x &&\in \Z_q.&
\end{align*}

By the winning condition in Game~\ref{game:dlog}, 
we have that $c_0 y_0/s_0 - c_1 y_1/s_1 \neq 0$,
which implies that we can compute:
\begin{align*}
  x &= \left(\frac{H(m_0)}{s_0} - \frac{H(m_1)}{s_1}\right) \cdot  \left (\frac{c_1 y_1}{s_1} - \frac{c_0 y_0}{s_0} \right)^{-1} \quad \in \Z_q.
\end{align*}
The algorithm $\Bdlog$ outputs this value $x \in \Z_q$.

\itpara{Case II: Adversary makes an $f$-oracle query.}
By the hypothesis of the lemma, we know that
the adversary makes a useful $f$-oracle query, which implies
that the adversary makes a single query to $f$ at the 
point $R = (g^{H(m_0)} X^{c_0 y_0})^{1/s_0} \in \G$.
Furthermore, since this condition holds for both forks of the execution, 
we have that 
\begin{align*}
  R = (g^{H(m_0)} X^{c_0 y_0})^{1/s_0} \quad \text{and}\quad
  R = (g^{H(m_1)} X^{c_1 y_1})^{1/s_1},
\end{align*}
where $(g,X) \in \G^2$ is the discrete-log challenge.
(Notice that the inverses of $s_0$ and $s_1$ in $\Z_q$ are well defined
because they are components of valid ECDSA signatures.) 

We now have a relation 
\begin{align*}
  \qquad(g^{H(m_0)} X^{c_0 y_0})^{1/s_0} &= (g^{H(m_1)} X^{c_1 y_1})^{1/s_1} &&\in \G&
\end{align*}
and, as in Case I, the winning condition in \gref{game:dlog} implies
that we can solve this linear relation in the exponent for the
discrete log $x \in \Z_q$.
\end{proof}

We now argue that the failure event in \gref{game:dlog} of Lemma~\ref{lemma:uf-single}
cannot happen often, as long as we instantiate the VIF construction with 
a secure VRF.

\begin{lemma} \label{lemma:vrf}
Let $\Wgame{game:fork}$ and $\Wgame{game:dlog}$ be the events defined
in the proof of Lemma~\ref{lemma:uf-single}.
Then there exists an algorithm $\Bvrf$ attacking 
VRF $\calV$ making $2\cdot \Qid$ VRF queries such that
  \[ \abs{\Pr[\Wgame{game:fork}] - \Pr[\Wgame{game:dlog}]} \leq 
\Qid \cdot \VRFAdv[\Bvrf, \calV] + \frac{2 \cdot Q_H}{q}.\]
Furthermore, $\Bvrf$ runs in time linear in the running time of $\calA$.
\end{lemma}

\begin{proof}
Let $F$ be the failure event that the adversary $\calA$ 
wins in \gref{game:fork} and outputs
forgeries that cause (\ref{eq:goodrel}) to fail to hold.
By the Difference Lemma~\cite{TheBook}, we have that 
$\abs{\Pr[\Wgame{game:fork}] - \Pr[\Wgame{game:dlog}]} \leq \Pr[F]$,
so we need only bound $\Pr[F]$.

Our strategy is to construct an efficient VRF set-targeting adversary $\Bvrf$ that:
  \begin{itemize}
    \item makes at most $2\cdot \Qid$ VRF queries,
    \item outputs a target set of size at most $Q_H$,
    \item runs in time linear in the running time of $\calA$, and
    \item satisfies $\Pr[F] \leq \VRFSetAdv[\Bvrf,\calV]$.
  \end{itemize}
Then, by applying Lemma~\ref{lemma:vrfset}, we can construct 
a VRF pseudorandomness adversary $\Bvrf'$ satisfying the conclusion
of the lemma such that 
\begin{align*}
\Pr[F] &\leq \VRFAdv[\Bvrf', \calV] + \frac{Q_H}{|\Z_q^*|}.
\end{align*}
Since $Q_H/|\Z^*_q| < 2\cdot Q_H/q$, this
proves the lemma.

Thus, our task is just to construct the VRF set-targeting adversary $\Bvrf$.
The algorithm $\Bvrf$ executes the following steps:
\begin{itemize}
  \item Receive the VRF public key $\pkVRF$ from the VRF set-targeting challenger.
        Generate a random $x \getsr \Z_q$ and send
        the pair $\mpk \gets (g^x, \pkVRF)$ to $\calA$ 
        as the VIF master public key.

  \item Run the algorithm $\calA$ until it either makes a
        signing query or outputs a forgery.
        \begin{itemize}
          \item Answer $\calA$'s signing-oracle queries by producing
                signatures using the master ECDSA secret key $x$. 
          \item Answer $\calA$'s identity queries by forwarding these
                to the VRF set-targeting challenger.
          \item Answer $\calA$'s $f$-oracle and $H$-oracle queries
                by defining these values lazily using a table.
        \end{itemize}

  \item At some point, $\calA$ either makes a signing query or outputs
        a forgery.
        In either case, we have a valid identity-message-signature 
        triple $(\id_0, m_0, (c_0, s_0))$.
        Compute:
        \[ r \gets (H(m_0) + x c_0 y_0)/s_0 \quad \in \Z_q, \]
        where $x$ is $\Bvrf$'s master ECDSA secret key and
        $y_0 \gets \VRFEval(\skVRF, \id_0)$, where $\skVRF$ is the challenger's
        VRF secret key.
        (Since the adversary is in standard form, as in Definition~\ref{defn:standard},
        the adversary queries the VRF 
        on $\id_0$ before outputting a forgery using $\id_0$ or 
        making a signing query on $\id_0$, algorithm $\Bvrf$ already 
        knows the value $y_0$ without needing to make an extra VRF query.)

        If the adversary made a signing-oracle query, let $c_1 \gets c_0$.
        If the adversary made an $f$-oracle query, 
        let $c_1$ be the value of the $f$-oracle returned 
        to the adversary in the right-fork.

  \item Let $h_1, h_2, \dots, h_{Q_H}$ be the values returned by the adversary's
        $H$-queries.
        (If the adversary has not yet made all $Q_H$ of its queries, 
        choose the undefined values at random.)

        For each such value $h_i$, compute the VRF output value $\hat{y}_i$ that,
        when used as part of the adversary's forged signature, would
        cause relation (\ref{eq:goodrel}) to fail to hold.
        That is, we solve the following two equations for $\hat{y}_i$:
        \[ \frac{c_0 y_0}{s_0} = \frac{c_1 \hat{y}_i}{s_1}\in\Z_q \quad \text{and} \quad s_1 = (h_i + xc_1 \hat{y}_i)/r \in \Z_q.\]
        We get:
        \[ \hat{y}_i = \frac{h_i c_0 y_0}{c_1 \cdot (s_0 r - c_0 y_0 x) } \in \Z_q. \]
        We must argue that the denominator of $\hat{y}_i$ is non-zero.
        Since $c_1$ is a component of a valid ECDSA signature, $c_1 \neq 0$.
        Then we must argue that $s_0 r - c_0 y_0 x \neq 0$.
        Since $s_0 = (H(m_0) + c_0 y_0 x)/r$, we have that
        $s_0 r - c_0 y_0 x = H(m_0)$.
        By the winning condition of \gref{game:zero} of Lemma~\ref{lemma:uf-single}, we 
        have that $H(m_0) \neq 0$. 
        Therefore, $\hat{y}_i$ is well defined.
        
        Finally, let $\calY^* = \{ \hat{y}_i \mid i \in \{1, \dots, Q_H\} \}$.

  \item Send the set $\calY^*$ to the VRF set-targeting challenger as the challenge set.
        Observe that the set $\calY^*$ has size at most $Q_H$.

  \item If the adversary made a signing-oracle query, continue executing the
        adversary.
        If the adversary made an $f$-oracle query, rewind the adversary
        as in \gref{game:fork} of Lemma~\ref{lemma:uf-single},
        reprogram the $f$ oracle, and continue executing the adversary. 

  \item When the adversary outputs a forgery $(\id_1, m_1, (c_1, s_1))$, 
        send the value $\id_1$ to the VRF set-targeting challenger.
\end{itemize}

For each identity query that $\calA$ makes, $\Bvrf$ makes a VRF query.
Since $\Bvrf$ could run $\calA$ twice (due to rewinding), the maximum
number of VRF queries that $\Bvrf$ makes is $2 \cdot \Qid$.

Now we must argue that $\Bvrf$ achieves the claimed advantage.
Whenever the failure event $F$ occurs, we have that the adversary $\calA$
gives us two valid signatures that fail to satisfy (\ref{eq:goodrel}).
Let the second forgery be $(\id_1, m_1, (c_1, s_1))$.
Let $\hat{y}^* \gets \VRFEval(\skVRF, \id_1)$, where $\skVRF$ is the VRF set-targeting
challenger's secret VRF key.
Then the fact that (\ref{eq:goodrel}) fails to hold implies that:
$c_0y_0/s_0 = c_1 \hat{y}^*/s_1 \in \Z_q$.

Since the adversary's forgery is valid, it holds
that $s_1 = (H(m_1) + c_1 \hat{y}^* x_1)/r \in \Z_q$.
Then, since $H(m_1)$ is in the set $\{h_1, \dots, h_{Q_H}\}$,
we know that $\hat{y}^* \in \calY^*$.
Thus, whenever event $F$ occurs, the output of $\Bvrf$ hits the set $\calY^*$.
\end{proof}

Finally, we show that even if the adversary can make many signing-oracle and
$f$-oracle queries, it still cannot break the VIF unforgeability property.
\begin{lemma}\label{lemma:many}
For any adversary $\calA$ attacking $\Fdsa$, defined over
a group $\G = \langle g \rangle$ of prime order $q$ using 
  \begin{itemize}
    \item $Q_f$ $f$-queries and 
    \item $\Qsig$ signing-oracle queries, 
  \end{itemize}
there is an adversary $\calB$ attacking $\Fdsa$ that
\begin{itemize}
  \item makes \emph{either} one signing-oracle query \emph{or} one $f$-oracle query and
  \item makes a useful query, in the sense of Definition~\ref{defn:useful}
\end{itemize}
such that:
\begin{multline*}
  \UFAdv[\calA, \Fdsa] \leq \\
  (Q_f + \Qsig) \cdot \UFAdvs[\calB, \Fdsa] \\+ \frac{2(Q_f + \Qsig)^2 + 1}{q}. 
\end{multline*}
Furthermore, the algorithm $\calB$ runs
in time linear in the running time of $\calA$.
\end{lemma}

\begin{proof}
We prove the lemma using a sequence of games.
We let $W_i$ denote the event
that the adversary ``wins'' in Game $i$.

\setcounter{GameCounter}{-1}

\game \label{game:manystart}
In this game, the adversary interacts with the VIF unforgeability challenger.
We say that the adversary wins in \gref{game:manystart} if the adversary
wins in the VIF unforgeability game.
Then, by definition,
\[ \Pr[\Wgame{game:manystart}] = \UFAdv[\calA, \Fdsa].\]

\game \label{game:fixup}
In this game, we change the way in which
the challenger replies to signing-oracle queries.
Namely, the challenger produces signatures without using
the secret ECDSA signing key by programming the $f$-oracle.
In particular, to reply to a signing query $(\id, m)$,
the challenger computes $y \gets \VRFEval(\skVRF, \id)$, 
chooses $c, s \getsr \Z^*_q$, sets
  \[ R \gets (g^{H(m)} X^{yc})^{1/s}\quad \in \G, \]
and defines $f(R) \deq c$.
If $f(R)$ is already defined, the challenger aborts.
As in \gref{game:start} of Lemma~\ref{lemma:uf-single}, we can
argue that the signatures produced in this way are statistically
indistinguishable to the adversary from signatures produced
using the secret ECDSA signing key.

Since Games~\ref{game:start} and~\ref{game:fixup} proceed
identically until there is an abort, to bound
  $\abs{\Pr[\Wgame{game:start}] - \Pr[\Wgame{game:fixup}]}$,
we need only bound the probability of an abort.

There are at most $Q_f + \Qsig$ points on $f$ defined,
and the probability that the challenger must redefine one such point
is at most $(Q_f + \Qsig)/(q-1) \leq 2(Q_f + \Qsig)/q$.
By a Union Bound over all queries, we get that the probability
  of ever aborting is at most $2(Q_f+\Qsig)^2/q$ and then
  \[ \abs{\Pr[\Wgame{game:start}] - \Pr[\Wgame{game:fixup}]} \leq \frac{2(Q_f + \Qsig)^2}{q}. \]

\paragraph{Final reduction.}
Finally, we construct an VIF unforgeability adversary $\calB$
that makes either 
a single signing-oracle query 
or a single useful $f$-oracle query 
and such that
  \[ \Pr[\Wgame{game:fixup}] = (Q_f + \Qsig) \cdot \UFAdv[\calB, \Fdsa]. \]

The algorithm $\calB$ interacts with the VIF unforgeability challenger
  and operates as follows:
  \begin{itemize}
    \item Make a guess $i^* \getsr \{1, \dots, (Q_f + \Qsig)\}$
          of which of the adversary's $f$- or signing-oracle queries will be useful,
          in the sense of Definition~\ref{defn:useful}.
          (Since the adversary is in standard form, as in Definition~\ref{defn:standard},
          at least one of these queries will be useful.)
    \item Upon receiving the VIF master public key $\mpk$, forward $\mpk$
          to algorithm $\calA$.
    \item As algorithm $\calA$ runs, it makes queries. 
          Respond to all of its queries as follows, except for the $i^*$-th 
          signing- or $f$-oracle query:
          \begin{itemize}
            \item Respond to signing-oracle queries by programming the $f$-oracle, 
                  as the \gref{game:fixup} challenger does.
            \item Forward identity-oracle queries to the VIF challenger
                  and return their responses to $\calA$.
            \item Respond to $f$- and $H$-oracle queries by building a 
                  table of responses, as the \gref{game:fixup} challenger does.
          \end{itemize}

    \item Upon receiving the $i^*$-th signing- or $f$-oracle query:
          \begin{itemize}
            \item If the $i^*$-th query is a signing-oracle query,
                  forward it to the VIF challenger and return
                  the reply to $\calA$.
                  (Notice that the response to the signing query fixes
                  a point in $f$. Algorithm $\calB$ must remember this 
                  point so that it responds consistently to $f$-oracle 
                  queries.)
            \item If the $i^*$-th query is an $f$-oracle query,
                  forward it to the VIF challenger
                  and return the reply to $\calA$.
          \end{itemize}
    \item Output whatever forgery the adversary $\calA$ outputs.
    \end{itemize}

The adversary $\calA$'s view in the interaction with $\calB$ is exactly 
as in \gref{game:fixup}, so the probability that $\calA$ outputs a 
forgery is exactly $\Pr[\Wgame{game:fixup}]$.

Furthermore, since the adversary is in standard form
(Definition~\ref{defn:standard}), the adversary $\calA$ always makes a useful
query (Definition~\ref{defn:useful}). 
With probability at least $1/(Q_f + \Qsig)$, the algorithm $\calB$
will correctly guess which of $\calA$'s queries is useful and thus
$\calA$'s forgery will be with respect to a point on 
the VIF challenger's $f$-oracle.
\end{proof}

\section{Security of key-generation protocol}
\label{app:keygen}

We now analyze the protocol of Section~\ref{sec:crypto:keygen}.

\paragraph{Bias-free (honest browser).}
If the browser is honest, then the values $v$ and $r$ it
chooses in Step~1 are distributed independently and uniformly
over $\Z_q$.
If the random oracle $H(\cdot,\cdot)$ yielded a perfectly
hiding commitment scheme, then 
the value $V' \in \G$ that the token sends 
in Step 2 would be independent of $v$ and thus the browser's output
$X = V' \cdot g^v \in \G$ would be uniform over $\G$.
Since the random oracle $H(\cdot, \cdot)$ only yields a
statistically hiding commitment, the resulting 
distribution is instead statistically close to uniform.

\paragraph{Bias-free (honest token).}
If the token is honest, then the value $v' \in \Z_q$ it
chooses in Step 2 of the protocol is distributed independently
and uniformly at random over $\Z_q$.
Once the browser commits to a pair $(v,r) \in \Z^2_q$ in Step 1, 
the probability that it can find a second pair $(v^*, r^*) \in \Z^2_q$
with $v \neq v^*$ such that $H(v,r) = H(v^*,r^*)$ is 
negligible. 
In other words, an efficient browser can only 
produce a single value $v$ in Step 3, except with negligible probability.
As long as this failure event does not occur, $v$ is distributed
independently of $v'$ and $x = v+v' \in \Z_q$ is distributed
uniformly over $\Z_q$.

\paragraph{Zero knowledge.}
Given an efficient adversary $\calA = (\calA_1, \calA_2)$ representing 
a malicious browser, we construct an 
efficient simulator $\Sim$ (Figure~\ref{fig:simzk}) that takes
as input a value $X \in \G$ and simulates the browser's view of the protocol.
The simulator need only work for an adversary that never
causes an honest token to output ``$\bot$.''
\begin{figure}
\centering
\begin{framed}
\begin{tabbing}
 \= 0 \= 00000000000000 \= \kill
\>$\Sim(X):$\\
\>\>$(\st, c) \gets \calA_1()$\\
\>\>$R \getsr \G$.\\
\>\>$(v,r) \gets \calA_2(\st, R)$.\\
\>\>$V' \gets (X/g^v) \in \G$.\\
\>\>$(v^*,r^*) \gets \calA_2(\st, V')$.\\
\>\>If $v \neq v^*$, abort.\\
\>\>Else, return $(c, V', (v,r^*))$.
\end{tabbing}
\vspace{-12pt}
\end{framed}
\caption{Simulator proving that the key-generation 
  protocol is zero knowledge.}\label{fig:simzk}
\end{figure}
To argue that the simulation is correct:
Since the adversary $\calA$ never causes the honest token to
output ``$\bot$,'' we know that $c = H(v,r) = H(v^*, r^*)$.
The binding property of $H$
(when viewed as a commitment scheme) implies that 
the probability that the simulation aborts due to
the $v \neq v^*$ event, is negligible.

Given that the simulation does not abort, the adversary's
view is simulated perfectly: the value $V'$ satisfies
$X = V' \cdot g^v$, as in the real interaction,
and the other values are identical as well.

\section{Security analysis of our 
firewalled signatures}
\label{app:san}

\begin{proof}[Proof sketch for Theorem~\ref{thm:san}]
We sketch the proof of Theorem~\ref{thm:san}. 

\paragraph{Exfiltration resistance.}
The bias-resistance of the key-generation protocol implies that
the value $R \in \G$ that $\calF$ holds at the end of Step~\ref{proto:san:1}
of the signing protocol is distributed statistically close 
to uniform over $\G$.
If $R = g^r$, then the value $\wR = g^{\pm r}$ is distributed exactly 
in the real ECDSA signing algorithm.

Having fixed $\wR$, there are only two possible valid signatures that $\calS^*$ can
output that are consistent with this choice of $\wR$: $\sigma$ and $\bar \sigma$.
In our protocol, $\calF$ outputs one of these two at random.
(Since $\calS^*$ never causes $\calF$ to abort, $\calS^*$ must have
sent a valid signature.)
The signatures output by $\ESign$ are distributed identically, since 
the ECDSA signing algorithm outputs $\sigma$ or $\bar \sigma$, depending on 
the ``sign'' of $r$ modulo~$q$.

\paragraph{Zero knowledge.}
Given a firewall $\calF^*$,
the simulator $\Sim(\pk, m, \sigma)$ operates as follows:
\begin{itemize}
  \item Use $\pk$, $m$, and $\sigma$ to solve for $\wR = g^{\pm r} \in \G$
        using Equation~(\ref{eq:ecdsa}) of Appendix~\ref{app:ecdsa}.
  \item Choose $R \getsr \{\wR, 1/\wR\}$ at random.
  \item Use the simulator for the key-generation protocol
        (given in the proof of its zero-knowledge property) on input $R$, with
        $\calF^*$
        playing the role of the browser, to get a transcript $\tau$ of
        $\calF^*$'s view in the key-generation protocol.
        If the simulated protocol aborts output $\tau$.
        Otherwise, output $(\tau, \sigma)$. 
        To invoke the simulator, we use the fact that $\calF^*$ never 
        causes the honest signer to abort.
\end{itemize}
To show that the simulation is correct, we argue that
(a) the simulation of the key-generation protocol is correct, using
an argument similar to the ZK argument for the key-generation protocol and 
(b) if the key-generation protocol is bias-free then $R$ is distributed
uniformly over $\G$ (as in the simulation). 
Thus, the entire simulation is correct.
\end{proof}

\end{document}